%% file: sample-sigconf-authordraft.tex
\documentclass[sigconf]{acmart}
\usepackage[most]{tcolorbox}
\usepackage{xcolor}
\usepackage[T1]{fontenc}
 \usepackage{subcaption}
 \usepackage{epigraph,multicol}
\usepackage{makecell, longtable}

\AtBeginDocument{%
  }

\begin{document}
\newcommand{\tanusree}[1]{{\color{green} \textbf{(Tanusree: #1)}}}
\title{Future of Algorithmic Organization: Large Scale Analysis of Decentralized
Autonomous Organizations (DAOs)}

\author{Tanusree Sharma}
\email{tfs5747@psu.edu}
\orcid{0000-0003-1523-163X}
\affiliation{%
  \institution{Pennsylvania State University}
  \city{State College}
  \state{PA}
  \country{USA}
}

\author{Yujin Potter}
\affiliation{%
  \institution{University of California, Berkeley}
  \city{Berkeley}
  \country{USA}}

\author{Kornrapat Pongmala}
\affiliation{%
  \institution{University of California, Berkeley}
  \city{Berkeley}
  \country{USA}
}

\author{Henry Wang}
\affiliation{%
 \institution{University of Illinois at Urbana-Champaign }
 \city{Champaign}
 \state{IL}
 \country{USA}}

\author{Andrew Miller}
\affiliation{%
\institution{University of Illinois at Urbana-Champaign }
 \city{Champaign}
 \state{IL}
 \country{USA}}

\author{Dawn Song}
\affiliation{%
 \institution{University of California, Berkeley}
  \city{Berkeley}
  \country{USA}}

\author{Yang Wang}
\affiliation{%
  \institution{University of Illinois at Urbana-Champaign }
 \city{Champaign}
 \state{IL}
 \country{USA}}


\renewcommand{\shortauthors}{Sharma et al.}

\begin{abstract}
Decentralized Autonomous Organizations (DAOs) resemble early online communities, particularly those centered around open-source projects, and present a potential empirical framework for complex social-
computing systems by encoding governance rules within ``smart contracts'' on the blockchain. A key function of a DAO is collective decision-making, typically carried out through a series of proposals where members vote on organizational events using governance tokens, signifying relative influence within the DAO. In just a few years, the deployment of DAOs surged with a total treasury of \$24.5 billion and 11.1M governance token holders collectively managing decisions across over 13,000 DAOs as of 2024.
In this study, we examine the operational dynamics of 100 DAOs, like pleasrdao, lexdao, lootdao, optimism collective, uniswap, etc. With large-scale empirical analysis of a diverse set of DAO categories and smart contracts and by leveraging on-chain (e.g., voting results) and off-chain data, we examine factors such as voting power, participation, and DAO characteristics dictating the level of decentralization, thus, the efficiency of management structures. As such, our study highlights that increased grassroots participation correlates with higher decentralization in a DAO, and lower variance in voting power within a DAO correlates with a higher level of decentralization, as consistently measured by Gini metrics. These insights closely align with key topics in political science, such as the allocation of power in decision-making and the effects of various governance models. We conclude by discussing the implications for researchers, and practitioners, emphasizing how these factors can inform the design of democratic governance systems in emerging applications that require active engagement from stakeholders in decision-making.

\end{abstract}





\maketitle

\input{sections/Introduction}
\input{sections/RelatedWork}
\input{sections/Method}
\input{sections/Results}

\input{sections/Discussion}

\onecolumn
\begin{multicols}{2}
\bibliographystyle{ACM-Reference-Format}
\bibliography{sample-base}
\end{multicols}
\appendix

\section{Analysis of Average Decentralization Level Based on Different Factor}


\begin{longtable}[t]{c||cccccc}

\caption{Evaluation of decentralization levels through correlation analysis and three metrics: Entropy, Gini coefficient, and Nakamoto.} \label{tab:voting_metric-1} \\

    \hline
    Name & Corr (P-value) & Entropy & Entropy-P & Gini & Gini-P & Nakamoto \\
    \hline
    \endfirsthead
    \caption[]{Evaluation of decentralization levels through correlation analysis and three metrics: Entropy, Gini coefficient, and Nakamoto (continued).} \\
    \hline
    Name & Corr (P-value) & Entropy & Entropy-P & Gini & Gini-P & Nakamoto \\
    \hline
    \endhead
    \hline
    \multicolumn{7}{r}{\textit{Continued on next page}} \\
    \endfoot
    \endlastfoot
    \hline\hline
1inch.eth  & 0.014 (0.627)& 5.189& 4.272&0.974&0.985&8\\
aavegotch.eth  & 0.124 (0.000)& 9.310& 7.688&0.936&0.973&105\\
aave.eth  & 0.012 (0.003)& 6.136& 4.008&0.999&1.000&17\\
arbitrumfoundation.eth  & 0.003 (0.584)& 5.653& 5.741&0.985&0.986&5\\
balancer.eth  & 0.065 (0.000)& 6.049& 3.896 & 0.991 &0.998 & 9\\
beetsdao.eth  & -0.168 (0.149)& 5.274& 5.082&0.603&0.651&10\\
cabindao.eth  & 0.628 (0.000)& 5.381 & 3.263 &0.803&0.942&8\\
cakevote.eth  & 0.013 (0.000)& 9.049 & 6.376 &0.989&0.996&45\\
club.eth  & 0.046 (0.486)& 5.573 & 5.477 & 0.830 & 0.848 & 10\\
council.graphprotocol.eth  & NA & 4.087 & 3.561 & 0 & 0.467 & 8\\
covalenthq.eth  & -0.027 (0.638) & 1.163 & 1.571 & 0.986 & 0.982 & 1\\
cream-finance.eth  & 0.086 (0.000) & 4.394 & 3.574 & 0.994 & 0.996 & 4\\
curve.eth  & -0.020 (0.610) & 3.880 & 3.510 & 0.968 & 0.973 & 2\\
daocity.eth  & -0.002 (0.902) & 11.293 & 10.607 & 0.253 & 0.610 & 1046\\
daoofdiamonds.eth  & 0.159 (0.0430) & 5.978 & 5.485 & 0.700 & 0.769 & 14\\
dappradar.eth  & 0.097 (0.002) & 6.231 & 5.776 & 0.922 & 0.948 & 15\\
datacrunch.eth  & 0.386 (0.000) & 5.567 & 4.880 & 0.733 & 0.826 & 11\\
devdao.eth  & 0.101 (0.000) & 6.086 & 5.467 & 0.969 & 0.980 & 10\\
dydxgov.eth  & -0.004 (0.780) & 4.277 & 4.386 & 0.994 & 0.995 & 4\\
elfork.eth  & 0.769 (0.000) & 5.364 & 5.075 & 0.412 & 0.562 & 18\\
fei.eth  & 0.056 (0.000) & 6.844 & 5.775 & 0.973 & 0.989 & 19\\
ffdao.eth  & 0.292 (0.000) & 5.658 & 4.471 & 0.829 & 0.919 & 9\\
freerossdao.eth & 0.034 (0.357) & 5.200 & 4.945 & 0.927 & 0.943 & 4 \\
friendswithbenefits.eth & 0.057 (0.006) & 9.587 & 8.780 & 0.611 & 0.795 & 154 \\
gaiadao.eth & 0.060 (0.625) & 4.856 & 4.706 & 0.685 & 0.715 & 7 \\
gasdao.eth & 0.085 (0.013) & 6.530 & 5.610 & 0.886 & 0.927 & 15 \\
gdao.eth & 0.070 (0.441) & 5.175 & 4.760 & 0.756 & 0.809 & 9 \\
genomesdao.eth & -0.006 (0.967) & 3.747 & 3.713 & 0.789 & 0.798 & 3 \\
gitcoindao.eth & 0.081 (0.000) & 5.986 & 4.875 & 0.993 & 0.997 & 14 \\
gmx.eth & 0.055 (0.000) & 7.453 & 6.987 & 0.996 & 0.997 & 29 \\
gnosis.eth & 0.010 (0.478) & 5.549 & 4.049 & 0.991 & 0.996 & 7 \\
goldhuntdao.eth & 0.416 (0.000) & 7.052 & 6.063 & 0.461 & 0.717 & 33 \\
golflinks.eth & 0.098 (0.000) & 10.413 & 9.848 & 0.464 & 0.648 & 450 \\
gov.radicle.eth & 0.285 (0.000) & 4.090 & 3.032 & 0.961 & 0.981 & 4 \\
humandao.eth & 0.024 (0.787) & 3.106 & 2.967 & 0.923 & 0.930 & 2 \\
instadapp-gov.eth & -0.023 (0.830) & 2.995 & 3.217 & 0.935 & 0.925 & 2 \\
insuredao.eth & 0.162 (0.234) & 3.437 & 3.233 & 0.848 & 0.870 & 3 \\
julswap.eth & -0.007 (0.638) & 3.353 & 3.465 & 0.995 & 0.995 & 2 \\
karastar.eth & -0.070 (0.248) & 4.896 & 4.263 & 0.908 & 0.918 & 7 \\
kcc.eth & 0.031 (0.158) & 9.028 & 8.596 & 0.791 & 0.826 & 185 \\
kleros.eth & 0.055 (0.332) & 5.617 & 5.527 & 0.865 & 0.882 & 9 \\
klimadao.eth & 0.015 (0.183) & 5.843 & 6.529 & 0.976 & 0.982 & 4 \\
kogecoin.eth & 0.103 (0.003) & 0.286 & 0.187 & 0.999 & 0.999 & 1 \\
kongsdao.eth & 0.409 (0.000) & 6.446 & 4.728 & 0.531 & 0.837 & 22 \\
leaguedao.eth & -0.234 (0.092) & 3.355 & 3.620 & 0.844 & 0.815 & 3 \\
lexdao.eth & -0.003 (0.982) & 6.277 & 5.493 & 0.021 & 0.564 & 39 \\
lido-snapshot.eth & 0.081 (0.000) & 5.654 & 4.119 & 0.993 & 0.998 & 12 \\
loopringdao.eth & 0.055 (0.001) & 8.986 & 8.518 & 0.849 & 0.882 & 88 \\
loot-dao.eth & 0.152 (0.000) & 8.591 & 7.380 & 0.638 & 0.806 & 70 \\
macaronswap.eth & 0.165 (0.001) & 6.899 & 6.283 & 0.764 & 0.826 & 23 \\
matrixdaoresearch.eth & NA & 6.129 & 5.981 & 0.000 & 0.240 & 35 \\
meebitsdao.eth & 0.299 (0.003) & 3.036 & 2.406 & 0.876 & 0.947 & 2 \\
nftx.eth & 0.390 (0.000) & 5.168 & 2.346 & 0.920 & 0.983 & 5 \\
officialoceandao.eth & 0.047 (0.203) & 6.277 & 6.062 & 0.889 & 0.922 & 11 \\
olympusdao.eth & 0.006 (0.582) & 7.624 & 7.799 & 0.957 & 0.966 & 21 \\
opcollective.eth & 0.033 (0.000) & 6.285 & 5.240 & 0.991 & 0.996 & 10 \\
partydao.eth & -0.153 (0.140) & 3.024 & 3.667 & 0.933 & 0.895 & 3 \\
people-dao.eth & -0.025 (0.191) & 6.110 & 4.906 & 0.963 & 0.975 & 9 \\
piedao.eth & 0.015 (0.653) & 5.102 & 6.017 & 0.946 & 0.939 & 6 \\
pleasrdao.eth & 0.429 (0.000) & 5.054 & 3.800 & 0.710 & 0.864 & 9 \\
polygonvalidators.eth & NA & 5.459 & 5.459 & 0.000 & 0.000 & 22 \\
polywrap.eth & 0.845 (0.000) & 4.121 & 2.211 & 0.763 & 0.931 & 4 \\
pooltogether.eth & 0.061 (0.059) & 6.303 & 5.821 & 0.931 & 0.949 & 16 \\
primexyz.eth & -0.196 (0.239) & 1.490 & 2.278 & 0.938 & 0.898 & 1 \\
raid-guild & 0.048 (0.821) & 3.560 & 3.430 & 0.620 & 0.668 & 3 \\
rarible.eth & -0.006 (0.785) & 4.554 & 5.121 & 0.975 & 0.977 & 2 \\
rbn.eth & -0.010 (0.711) & 7.714 & 7.790 & 0.802 & 0.829 & 25 \\
rook.eth & 0.066 (0.229) & 6.455 & 5.912 & 0.798 & 0.856 & 20 \\
shapeshiftdao.eth & 0.030 (0.191) & 6.068 & 5.796 & 0.963 & 0.976 & 12 \\
sharkdao.eth & 0.339 (0.000) & 7.045 & 4.389 & 0.820 & 0.958 & 25 \\
sismo.eth & 0.064 (0.000) & 9.801 & 8.771 & 0.758 & 0.869 & 129 \\
snapshot.dcl.eth & 0.033 (0.006) & 6.812 & 5.651 & 0.979 & 0.993 & 15 \\
songadao.eth & NA & 5.907 & 5.534 & 0.000 & 0.389 & 30 \\
stgdao.eth & 0.008 (0.004) & 4.823 & 4.697 & 0.989 & 0.995 & 4 \\
sushigov.eth & 0.013 (0.176) & 5.716 & 5.139 & 0.994 & 0.996 & 9 \\
tempusgov.eth & -0.067 (0.469) & 4.356 & 4.662 & 0.853 & 0.832 & 4 \\
temp.metricsdao.eth & -0.152 (0.376) & 5.155 & 4.793 & 0.026 & 0.398 & 18 \\
theplantdao.eth & 0.207 (0.138) & 5.016 & 4.922 & 0.511 & 0.551 & 12 \\
tomoondao.eth & 0.128 (0.000) & 8.785 & 7.725 & 0.857 & 0.915 & 88 \\
tosinshada.eth & NA & 3.585 & 3.585 & 0.000 & 0.000 & 6 \\
trashdao.szns.eth & 0.468 (0.016) & 3.342 & 2.247 & 0.691 & 0.832 & 3 \\
treasuregaming.eth & -0.033 (0.037) & 4.697 & 4.468 & 0.995 & 0.995 & 7 \\
truefigov.eth & 0.300 (0.000) & 4.642 & 3.860 & 0.901 & 0.943 & 5 \\
trustwallet & 0.074 (0.000) & 8.576 & 8.108 & 0.950 & 0.963 & 56 \\
ubi-voting.eth & 0.285 (0.000) & 3.058 & 2.514 & 0.940 & 0.972 & 2 \\
uniswap & -0.004 (0.527) & 5.509 & 4.890 & 0.999 & 0.999 & 13 \\
unlock-protocol.eth & -0.028 (0.722) & 2.623 & 3.845 & 0.948 & 0.928 & 1 \\
vote-perp.eth & 0.019 (0.512) & 5.175 & 4.575 & 0.973 & 0.981 & 8 \\
vote.vitadao.eth & 0.052 (0.346) & 4.654 & 4.770 & 0.929 & 0.936 & 5 \\
yam.eth & 0.144 (0.000) & 6.825 & 5.175 & 0.979 & 0.991 & 19 \\
yearn & 0.064 (0.000) & 6.714 & 6.873 & 0.961 & 0.964 & 15 \\
      \hline\hline
\makecell{Average \\(Std)} & \makecell{0.100 \\(0.186)} & \makecell{5.618 \\(1.989)} & \makecell{5.04\\ (1.834)} & \makecell{0.794 \\(0.275)} & \makecell{0.857\\ (0.202)} & \makecell{35.923\\ (120.432)} \\ \\
\label{tab:voting_metric-1}
\end{longtable}

\end{document}

%% file: sections/Introduction.tex
\section{Introduction}

Decentralized Autonomous Organizations resemble early online communities, especially those focused on open-source projects.  They also draw inspiration from various models, including digital and platform cooperatives~\cite{mannan2018fostering}, multi-organizational networks like keiretsus~\cite{lincoln1996keiretsu}, crowdfunding platforms such as Patreon, virtual economies in games like World of Warcraft and Second Life~\cite{lehdonvirta2014virtual} and peer-produced projects like Wikipedia~\cite{xu2015empirical}. DAOs challenge the traditional roles of firms by providing a potential empirical testbed for exploring social choice experiments in potentially improving the current governance structure as well as improving longstanding social, organizational, and legal issues through a computational lens~\cite{benkler2015peer}. As such, DAOs present unique opportunities to implement innovative mechanisms in social choice designs, including,  quadratic voting, and forms of liquid democracy~\cite{lalley2018quadratic, weyl2022decentralized, zhang2017brief}. DAO governance, as human-centric digital organizations, addresses key issues in social computing but is more complex than platforms like civic tech~\cite{poor2005mechanisms}, blockchain governance, online communities~\cite{love2010linux}. DAOs emphasize entrepreneurial functionality with less reliance on existing institutional hierarchy, with greater emphasis on legal, economic, and organizational sophistication. They prioritize governance over traditional management tools (e.g., Slack, Salesforce, Zendesk), which do not manage rights and power digitally. 




In just a few years, the deployment of DAOs surged with a total treasury of \$24.5 billion and 11.1M governance token holders collectively managing decisions across over 13,000 DAOs as of 2024~\footnote{\url{https://deepdao.io/organizations}}. Technologists argued that DAOs could automate many organizational processes and allow for more broad-based ownership and governance of the digital economy, all on the basis of a cryptographically secured blockchain~\cite{buterin2014daos}. DAOs transcend being mere novelties within online communities; they possess the transformative power to shift vast segments of institutional frameworks from the physical realm to the digital world, merging legal and economic principles with computer science. DAOs in their current form may or may not become the future of organizations, the increasing significance of online organizational forms in global politics and economy is undeniable. Research in DAOs presents a promising avenue for addressing complex issues related to organization, coordination, and governance, marking it as a critical area of study in the evolving landscape of digital and organizational sciences.

A key function of a DAO is collective decision-making, typically carried out through a series of proposals where members vote on organizational events using governance tokens, signifying relative influence within the DAO. The process of voting on proposals is central to the governance of DAOs. Voting raises many design decisions, such as how to allocate voting power and how voting choices should translate into outcomes~\cite{sharma2023unpacking}. These decisions relate to questions of social choice~\cite{shaw2002public} and tokenomics~\cite{sharma2022s}. DAO governance relies on secure and trustworthy voting mechanisms, which differ significantly from traditional political voting. First, DAO voting is entirely digital, eliminating the need for physical polling stations or ballots. Second, it is typically pseudonymous, with voters identified by on-chain addresses instead of real-world identities. Finally, unlike most political elections that allow one vote per eligible voter, DAO voting often employs various methods that weight votes based on token holdings. DAO voting bears some similarity in this sense to corporate proxy voting. In previous works, researchers highlighted social and technical metrics such as token distribution, voter participation, voting delegation, and voting behavior patterns~\cite{sharma2023unpacking, feichtinger2023hidden}. Some tension arises from social measures such as trust and community belonging, which influence decision-making and key functions within economic coordination such as, voting power distribution and equality, and level of decentralization~\cite{berg2017exit, sharma2023unpacking, fritsch2022analyzing, sharma2024can}.

Building on existing case studies of a limited number of DAOs to identify factors influencing their success or failure, as well as stakeholder interviews~\cite{sharma2023unpacking, feichtinger2023hidden}, this study aims to evaluate the significance of various governance components that enhance governance quality.
We conducted a large-scale empirical analysis of 100 DAOs with a diverse set of categories and smart contracts~\cite{sharma2023mixed} by leveraging on-chain (e.g., voting results, proposal data) and off-chain data.
We examine how governance mechanisms, such as, voting power, token design contribute to varying levels of decentralization across different DAOs. 
In particular, this paper addresses the following research questions.

\textbf{RQ1}: What factors influence the degree of decentralization in Decentralized Autonomous Organizations (DAOs)?

\textbf{RQ2}: How do voting mechanisms affect the degree of decentralization in DAOs?

\textbf{Main Findings.} 
Our analysis across various DAOs revealed statistically significant results between DAO properties—such as token market value, DAO type, and voting power distribution—and decentralization levels. We found social or public-good-focused DAOs (e.g., sports DAOs) showing lower Gini coefficients, indicating higher decentralization, while infrastructure and investment DAOs demonstrated greater centralization. DAOs with higher secondary market token values often had poor decentralization. We observed notable power imbalances in DAOs like 1inch, Balancer, dYdX, Fei, GMX, and Gnosis, with low voter participation and high centralization. In our analysis, we found Gini metrics to be consistent across DAOs, suggesting it can be leveraged by stakeholders to monitor decentralization trends over time. 

Our findings also highlight that the governance mechanisms, such as, power distribution, incentive design, and grassroots participation can inform design choices for governance in emerging areas such as AI and technology governance, where researchers and practitioners are actively exploring potential democratization among various stakeholders~\cite{openai}.

%% file: sections/RelatedWork.tex
\section{Related Work}
\label{background}
\subsection{Background of DAOs}
The concept of a DAO has existed since the mid-2010s~\cite{chohan2017decentralized} when DAOs were envisioned as digital alternatives to conventional organizations, promising automation of organizational processes and broader ownership and governance in the digital economy on the basis of a cryptographically secured blockchain~\cite{larimer2013bitcoin, buterin2014daos}. While DAOs with such operations existed mostly in the realm of speculation~\cite{dupont2017experiments}, the reality of DAOs has drastically changed over the past few years. While some researchers initially argue that DAOs were limited to private capital allocation~\cite{chohan2017decentralized, trisetyarso2019crypto}, such as, Digix.io\footnote{Digix.io is a smart-asset gold-focused coin.}, Augur\footnote{Augur centers around the prediction markets where financial options and insurance markets can be developed.}, Uniswap\footnote{Uniswap is a crypto exchange based on smart contracts}, there is a growing trend to use DAOs in high-value data, and reputational-based systems~\cite{myeong2019administrative, barbosa2018cryptocurrencies,chohan2017decentralized, wang2022empirical, messias2023understanding} and even the first DAO, \emph{The DAO}, was originally designed as an investor-driven venture capital fund that relied on voting by investors to disburse funds to proposals submitted by contractors and vetted by curators~\cite{mehar2019understanding}. It operated as a transparent and democratically structured virtual platform, without formal managerial roles. 

While the primary objective of a DAO is to replicate traditional organizations with distributed decision-making by using blockchain technology, the precise definition of a DAO is currently contested~\cite{dupont2017experiments}. Some argue that all cryptocurrencies can be considered DAOs, while others propose that decentralized and autonomous governance structures could be considered the original DAOs~\cite{morrison2020dao}. As such, the specific tasks that require coordination within a DAO and what it means to \emph{``coordinate''} in this context of blockchain require further exploration.


\vspace{-2mm}
\subsection{Co-ordination in Social Collaborative Systems}
\vspace{-1mm}

Governance, particularly coordination, plays a crucial role in project-based activities within online communities, open-source projects, and collective intelligence platforms such as Wikipedia~\cite{ung2010project}, Slashdot~\cite{poor2005mechanisms}, the Linux Kernel~\cite{love2010linux}, and Zooniverse~\cite{simpson2014zooniverse}. Previous research highlights the systematic benefits these communities offer compared to markets and hierarchical management structures in digitally networked environments, especially in information or cultural production~\cite{benkler2002coase}. The success of such communities depends on effective coordination, governance, trust, and the scalability of collaboration~\cite{kittur2009coordination}. While not always explicitly labeled as DAOs (Decentralized Autonomous Organizations), a wealth body of literature in areas like the theory of firms~\cite{williamson2002theory}, public choice theory~\cite{shaw2002public}, and platform cooperatives~\cite{ostrom1990governing, scholz2016ours} addresses themes such as decentralized governance, incentive alignment, self-management, group dynamics and paradoxes in decision making~\cite{nurmi1999voting}, themes central to the structure of DAOs. DAOs transcend being mere novelties within online communities; they possess the transformative power with technical solutions to shift vast segments of institutional frameworks, merging legal and economic principles with computer science.

Furthermore, coordination in management science is critical to ensure that resources are used efficiently and that organizations work towards the same objectives~\cite{faraj2006coordination}. Performance metrics and feedback mechanisms~\cite{faraj2006coordination}, as well as computer-based coordination tools~\cite{fish1988quilt, stokols2008ecology}, are deemed to be necessary to track progress towards organizational goals. Similarly, DAOs, as internet-native organizations, manage coordination digitally with a protocol specified in code and enforced on the blockchain.
When environmental change is high, organizational systems need to adapt quickly, and this work is typically facilitated by people who focus almost exclusively on coordination as opposed to execution and that is the role of management, or, put into the language of DAOs, that is the role of community managers and delegates~\cite{burton2017github}. DAOs represent a unique combination of the concepts of Autonomous Agents (without human involvement) and Decentralized Organizations (without external influence), blurring the boundaries between the two. 

Another important factor in online community is the influence of ones' voice in the community~\cite{liu2022user,guidi2020graph}, akin to the voting power in DAOs~\cite{li2019incentivized, guidi2020graph, guidi2021analysis, sharma2023unpacking, austgen2023dao, ma2024demystifying, pateman1970participation}. Recent studies suggest that governance tokens in DAOs can be vulnerable to bribery, affecting decision-making processes. This is comparable to both online and offline communities where influential users can manipulate recommendations. Bribery in these contexts can exploit social influence and create biased outcomes\cite{ramos2020negative}. Voting paradoxes and the Gibbard-Satterthwaite theorem~\cite{benoit2000gibbard} highlight that an individual's voting power can be influenced by the structure of the voting system and the distribution of preferences, and no system is perfect at consistently representing voter preferences~\cite{satterthwaite1975strategy}. Although prior research has explored the voting power of DAO holders on the Ethereum blockchain~\cite{fritsch2022analyzing}, it has not fully addressed the importance of participation, which is crucial for legitimizing decisions~\cite{pateman1970participation}. In this study, we aim to investigate the relationship between individual influence and participation, proposing design suggestions to encourage broader adoption of DAOs.

\vspace{-2mm}
\subsection{DAOs in Technological Governance}

Scientists and practitioners have been quick to acknowledge the potential power of DAOs in the context of Governance, and institutional economics, organizational imprinting~\cite{tan2023open, sharma2024experts, openai}. The emergence of Decentralized Autonomous Organizations (DAOs) introduces possible solutions for various challenges related to political institutions, including classic coordination dilemmas such as preference aggregation, credible commitments, audience costs, information asymmetry, representation, and accountability~\cite{hall1996political}. This unique empirical context offers an opportunity for political scientists to examine fundamental theories concerning political institutions and develop innovative theories that can be tested within digital governance. Political institutions encompass both formal and informal rules, procedures, and organizations governing individuals' and groups' behavior within a political system~\cite{hall1996political}. These institutions represent the \textit{rules of the game} or the constraints shaping human interaction~\cite{north2018institutional}. Scholars have engaged in debates about the consequences of different institutional designs, such as the separation of powers, federalism, the strategic configuration of non-democratic institutions, and processes of institutional change. The relevance of these theories to the design of digitally-native governance institutions is a critical question. For instance, the separation of powers in DAOs impacts the prevention of excessive concentration of power, enhances transparency and accountability, or potentially leads to governmental gridlock and indecision~\cite{de1989montesquieu}. Governance questions have puzzled societies and organizations, with the contested concept of democracy, a system contingent on the will of the people, occupying a central place in this debates~\cite{rousseau1964social, dahl1989democracy, landemore2012democratic}. This has gained new relevance with emerging technologies like AI, where inclusive and representative decision-making is critical throughout the development lifecycle. In theory, DAOs present innovative opportunities for collective decision-making. However, challenges persist, especially concerning technocracy, which continues to be addressed through adaptable mechanisms like quadratic funding~\cite{buterin2019flexible} and automated decision-execution protocols.

%% file: sections/Method.tex
\section{Method}

\subsection{Selection Criteria \& Method} We selected 100 DAOs, which were formed between 2019 and 2023, prioritizing diverse categories (e.g., investment, defi, social, etc), popularity, market capitalization, type of Treasury, and the nature of governance token. 

\textbf{DAO types.} DAO types mainly refer to the primary functionality and operational focus of the organization, both at the community and smart contract levels. For instance, Sushiswap and Lido DAO are DeFi protocols, while CityDAO and LinksDAO are categorized as social DAOs.

\textbf{Type of Treasury.} DAO treasury is the pool of funds for the continuous growth and development of the organization. Members of the DAO rely on the specified governance mechanisms for determining the allocation of the DAO treasury funds. 
We referenced DAOHQ\footnote{DAOHQ: \url{https://www.daohq.co/}}, DeepDAO\footnote{DeepDAO: \url{https://deepdao.io/organizations}}, popular publicly available analytics platform of DAOs, to determine the threshold of treasury among currently available DAOs. Generally, the threshold of the treasury is divided into four main categories, including (0-\$1m, \$1m - \$10m, \$10m-\$50m, \$50m+). We utilized this threshold to select DAOs with diverse treasury token holdings. We distributed 0-1m as low treasury, \$1-10m as medium, and \$10-50m+ as high treasury. 



\textbf{Nature of governance tokens (monetary vs. non-monetary).}  We used DAOs with no secondary market value as non-monetary where those types of DAOs largely fall under social, research DAOs. More specifically, governance tokens that cannot be sold in the secondary market, such as the reputation-based and nontransferable dxDAO token.

\textbf{Stratified Sampling.} We applied a stratified sampling method to select 100 DAOs from a pool of 2391 DAOs in DeepDAO analytics. We divided the population into different groups based on four factors. For each of the factors, DAOs are subdivided into different groups where DAOs in each group are similar to the other. Every stratum is sampled and therefore represented. 

The four factors, we utilized are (1) DAO Types, (2) DAO Treasury Amount, (3) DAO voting module, and (4) the Nature of the governance token. For each of the factors, there are subgroups where the DAOs are similar to each other. For example, for factor (1) DAO types 10 types of DAOs such as Defi protocol, investment, DAO tool, NFTs, research, etc. For factor (2) treasury size, our population has 4 strata, including DAO with $\$0-\$1m$, $\$1-\$10m$, $\$10m-\$100m$ and 100m+. For factor (3) DAO voting module, there are three strata, including (a) snapshot, (b) on-chain, and (c) Others. For factor (4) Nature of DAO tokens, there are two types.
By applying random stratified sampling, we select 100 DAOs. We separated subgroups (combination) for each criterion and randomly selected a proportionate number of DAOs from each subgroup. We finally evaluate the characteristics of the final sample and compare them to the overall population of 2391 DAOs to ensure that the selected DAOs sufficiently represent the three types of selection criteria and provide a diverse and representative sample for our analysis.

\vspace{-2mm}
\subsection{Data Collection}
\vspace{-2mm}
To address the research question, we identify and collect the necessary data by automating the process for voting data. 

Our objective is to test the influence of various factors on the level of polarization in DAOs, similarities/ differences in their voting behavior, and the degree of decentralization over time. These factors include the token holdings of DAO holders, tokens with higher secondary market value, the voting choices (for or against) made by members, voting power of DAO members, and participation. To accomplish this, we require a dataset that encompasses the voting data of DAO members, including attributes such as DAO holders' addresses, delegated voting power during voting for different proposals, timestamps, proposal IDs, and voting choices for different proposals.
The majority of the DAOs leveraged the snapshot platform for voting, while some utilized on-chain voting or other platforms such as the Aragon app or manual community consensus. Accordingly, we developed four types of voting data crawlers tailored to the different voting methods employed by the DAOs. For DAOs that operate completely on-chain (e.g., makerDAO), we utilized libraries like Web3.js or Ethers.js, along with the Python programming language, to query relevant contract methods and retrieve information about proposals, votes, voter details, timestamps, and other pertinent data points. For platforms like Snapshot, we utilized publicly available GraphQL APIs to crawl the voting data. 
 \vspace{-2mm}
\subsection{Data Analysis} 
 \vspace{-2mm}
\label{data-analysis-chapter7}
Drawing from the related work (section~\ref{background})
we formulate hypotheses. 
\begin{tcolorbox}[width=\linewidth, colback=white!95!black, boxrule=0.5pt, left=2pt,right=2pt,top=1pt,bottom=1pt]
  \emph{\textbf{H1: Governance Token}: Governance token with a higher secondary market value decreases the level of decentralization within a DAO.}

  \emph{\textbf{H2: DAO Type}: Different DAO types are more likely to influence the level of decentralization}
  
  \emph{\textbf{H3: Market Cap}: DAOs with a high market cap are more likely to have a lower decentralization}
  
\emph{\textbf{H4: Treasury}: DAOs with a high treasury are more likely to have a low decentralization level.}

\emph{\textbf{H5: Participation Rate}: An upward trend in grassroots participation is more likely to increase the trend in decentralization in a DAO} 

\emph{\textbf{H6: Voting Power}: A lower variance in voting power within a DAO is more likely to increase the level of decentralization in a DAO.}
\end{tcolorbox}

\textbf{Factors Impacting Level of Decentralization.} We conduct statistical analysis considering the following factors and decentralization levels using Gini, Nakamoto, and entropy metrics where predictors are subsequently, Marketcap, Treasury, Token secondary market, and DAO Type.

\textbf{Evaluation of Decentralization Levels with Voting Power.} We conducted a correlation analysis between voters’ token weight and a participation rate, which is defined as the number of proposals that a voter participated in. We aim to determine
whether the participation rate, therefore the voting power, is biased toward rich token holders. Here, the positive correlation between the participation rate and token weight implies poor decentralization due to the more active participation of only richer voters than their counterparts.
Next, we evaluated the decentralization level of 100 DAOs in terms of voting power. To do this, we used three metrics: Entropy, the Gini, and Nakamoto coefficient. Entropy can be generally used to assess randomness and degrees of freedom for a given system. Mathematically, the metric for given a series of numbers $x (={x_i})$ is defined as follows: 
$Entropy(x)$=$\sum_{i}log_2\left(\frac{x_i}{\sum_{i}x_i}\right)*\frac{x_i}{\sum_{i}x_i}.$
If the value of entropy is high, it implies a DAO has a high degree of randomness and freedom, which is related to a high level of decentralization. The Gini coefficient used to measure wealth inequality traditionally can be applied to assess the voting power inequality among voters in our analysis. The metric for given a series of numbers x is defined as follows: 
$Gini(x)$=$\frac{\sum_{i,j}|x_i-x_j|}{2 |x| \sum_{i}x_i},$
where $|x|$ indicates the length of the number series. The minimum value of the Gini coefficient as 0 would represent the full equality among voters, while the maximum value of 1 would represent the full power bias towards one voter. We calculated Entropy and Gini based on the token weight of voters and their voting scores, which, in this study, we define as \textit{``token weight$\times$the number of proposals that a voter participated in''} to reflect not only the token weight but also a participation rate. Lastly, the Nakamoto coefficient indicates the minimum number of entities that can subvert the system. In the voting context, the entities can accept or reject the proposal in accordance with their interests, even if it may be against most users. Therefore, if power is significantly biased towards only a few voters, the DAO cannot be considered to attain true democracy, which is one of the main goals of DAOs. We calculated the Nakamoto coefficient based on the token weight of voters.    
We also ran a series of linear regression analyses, and logistic regression regressing the level of decentralization outcome of each DAO on the respective set of characteristics and controls, including secondary market value, market cap, DAO types, treasury, and voting power allocated for different proposals. 

%% file: sections/Results.tex
\section{Result}
\begin{figure*}[t!]
    \centering
    \begin{subfigure}[t]{0.23\textwidth}
        \centering
        \includegraphics[width=\linewidth]{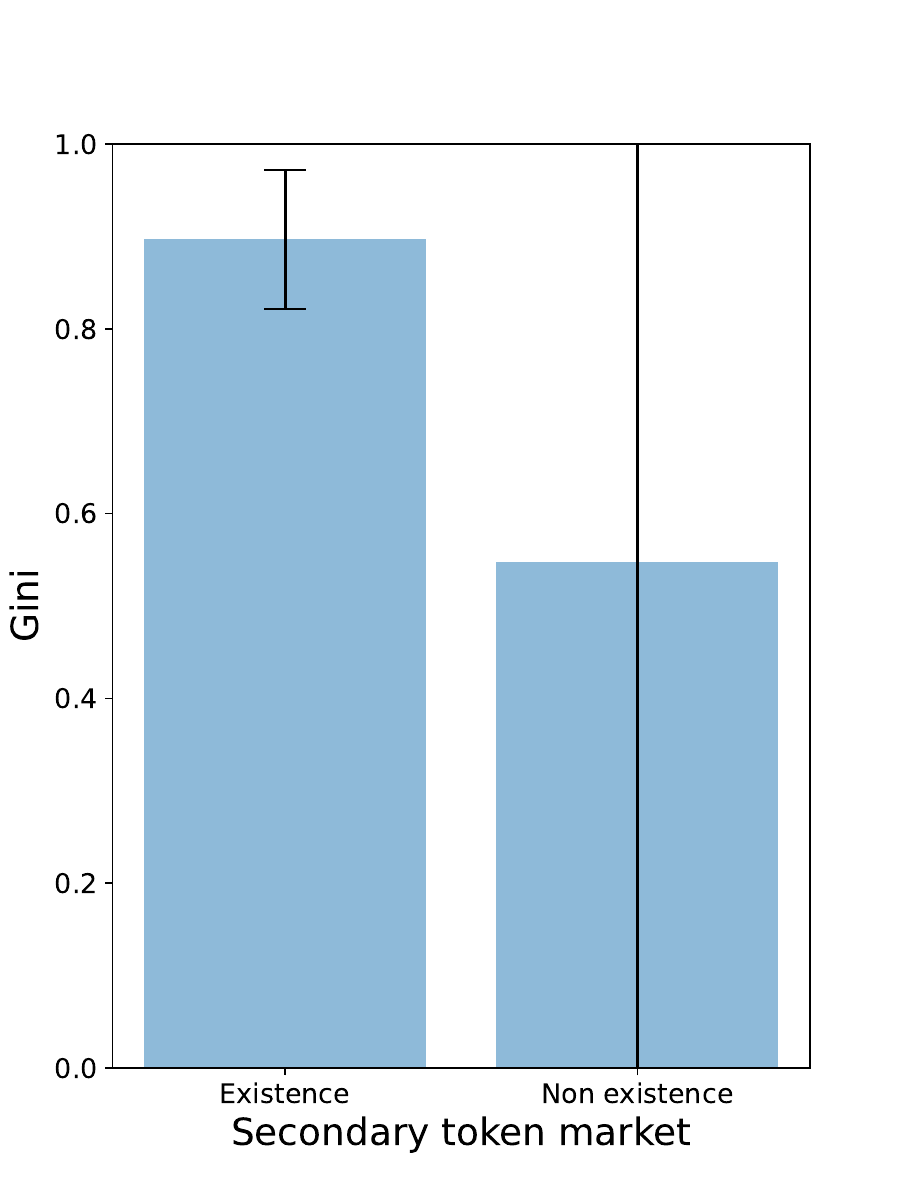}
        \caption{average Gini vs secondary market value}
        \label{fig:registration}
    \end{subfigure}%
    ~~
    \begin{subfigure}[t]{0.23\textwidth}
        \centering
        \includegraphics[width=\linewidth]{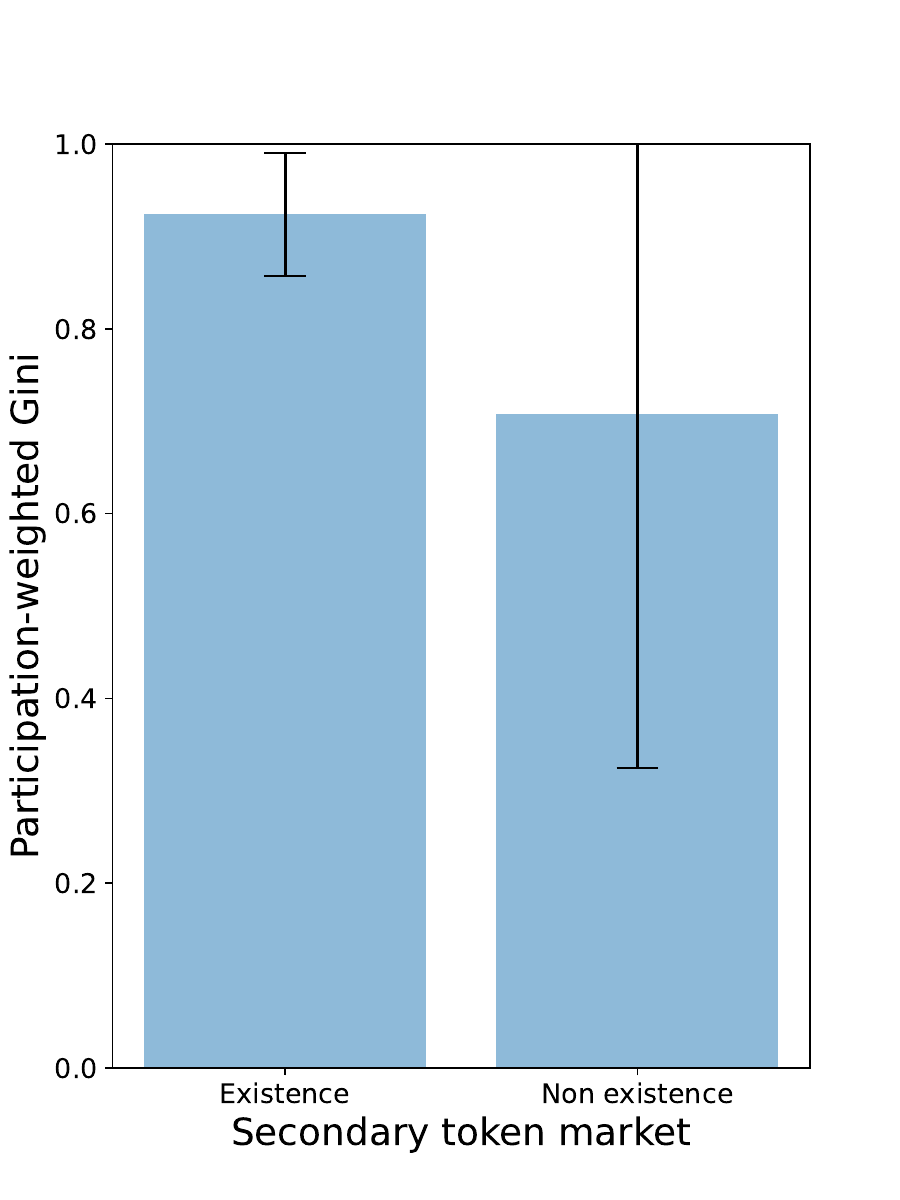}
        \caption{average participation Gini vs secondary market value}
        \label{fig:login}
    \end{subfigure}%
    ~~
    \begin{subfigure}[t]{0.23\textwidth}
        \centering
        \includegraphics[width=\linewidth]{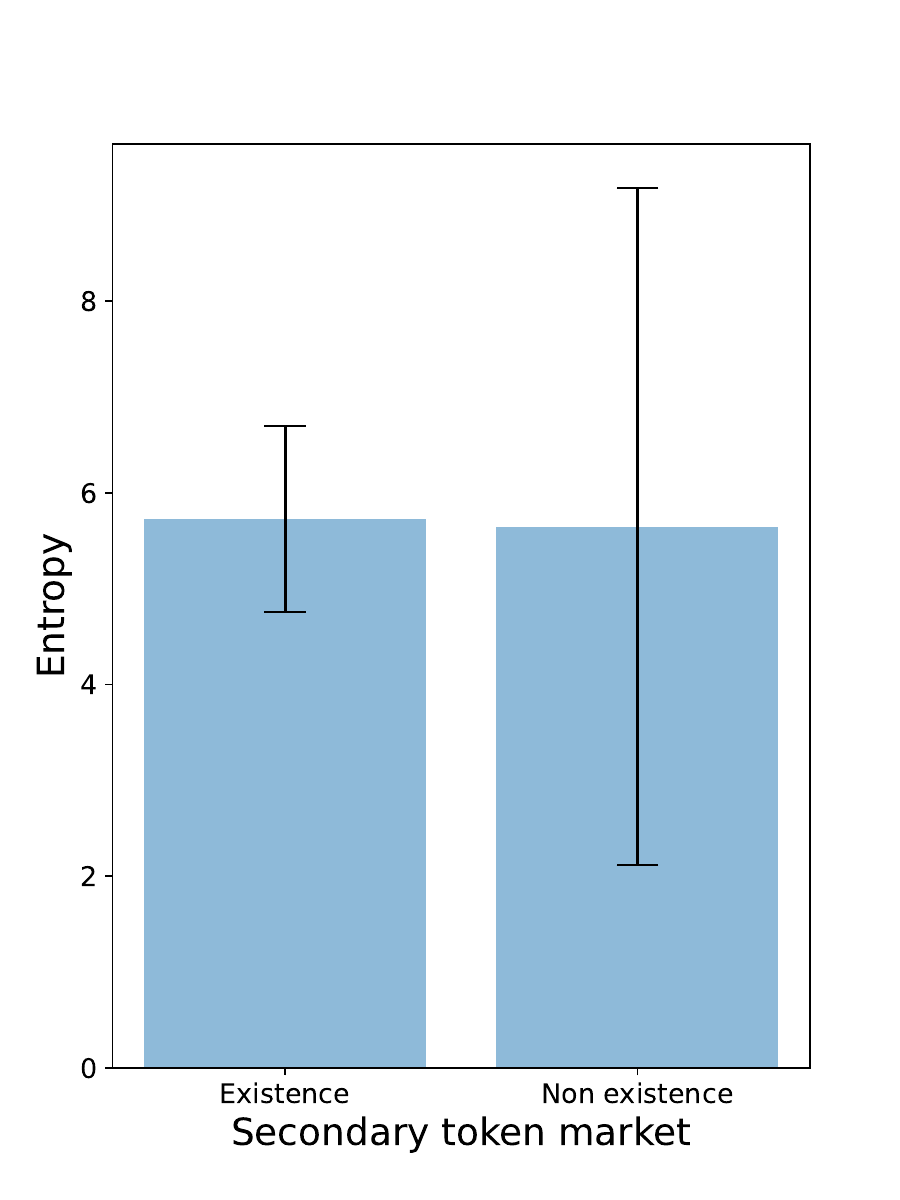}
        \caption{average entropy vs secondary market value}
        \label{fig:recovery}
    \end{subfigure}
    ~~
    \begin{subfigure}[t]{0.23\textwidth}
        \centering
        \includegraphics[width=\linewidth]{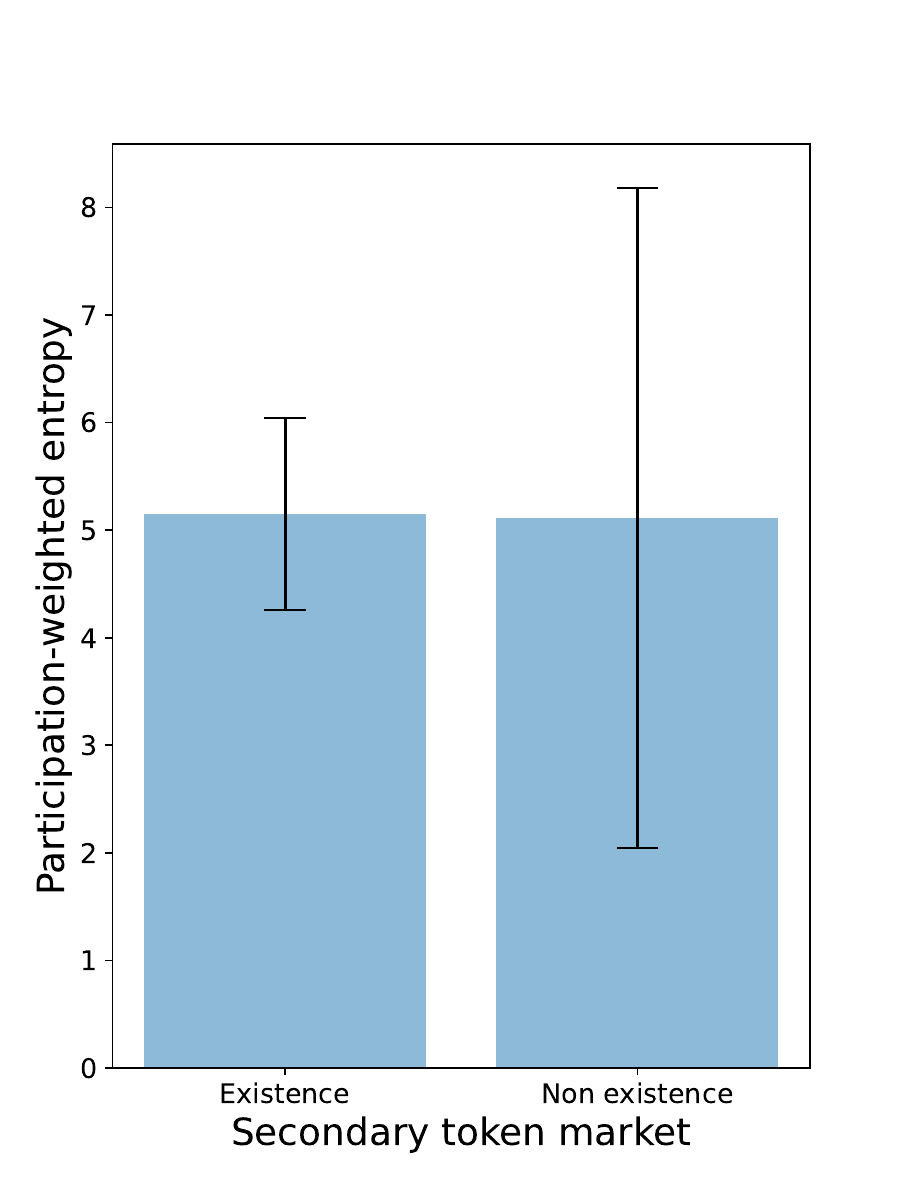}
        \caption{average participation entropy vs secondary market value}
        \label{fig:recovery}
    \end{subfigure}
    \caption{Average decentralization level (Secondary Market Value) with Decentralization Metrics
    }
    \label{fig:secondary}
\end{figure*}
\begin{figure*}[t!]
    \centering
    \begin{subfigure}[t]{0.30\textwidth}
        \centering
        \includegraphics[width=\linewidth]{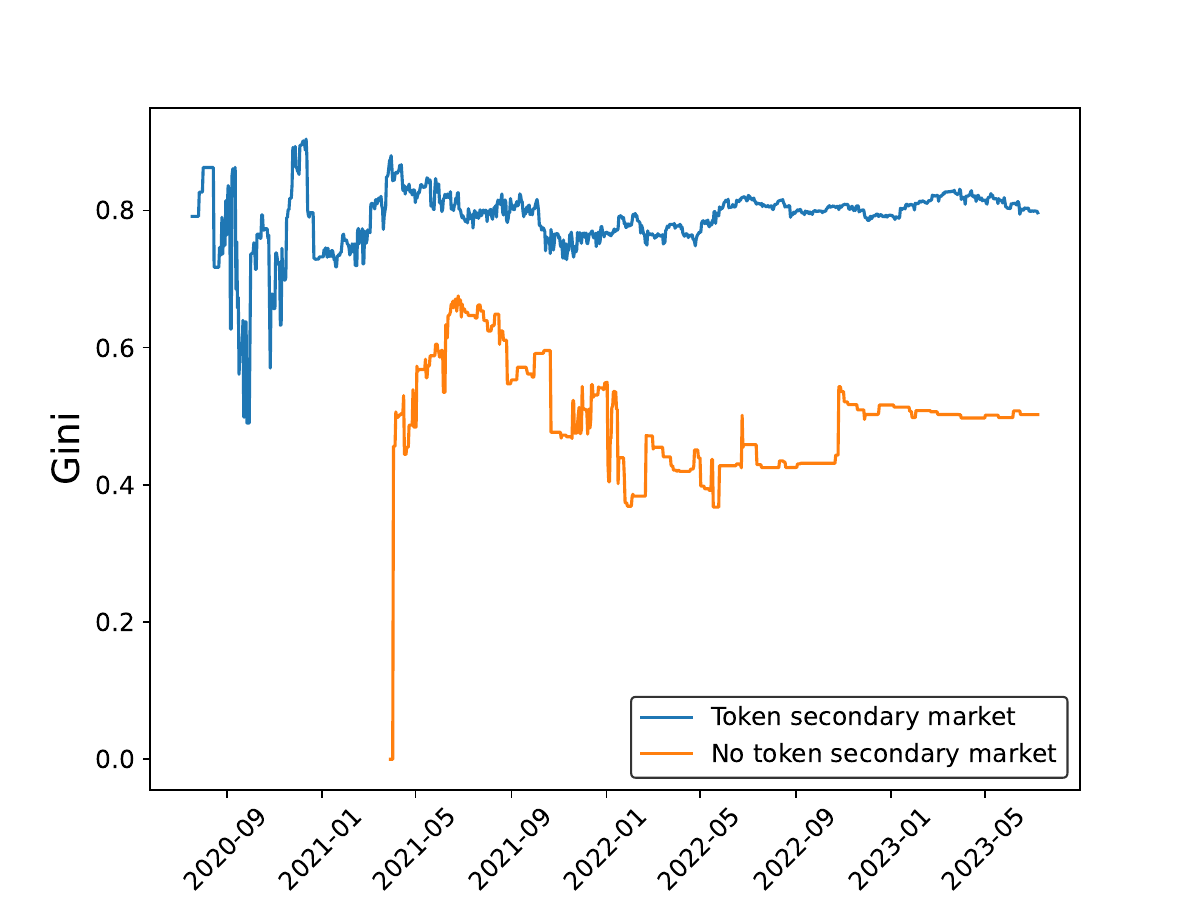}
        \caption{Demonstration with Gini index vs DAO with Secondary Market Value}
        \label{fig:registration}
    \end{subfigure}%
    ~~
    \begin{subfigure}[t]{0.30\textwidth}
        \centering
        \includegraphics[width=\linewidth]{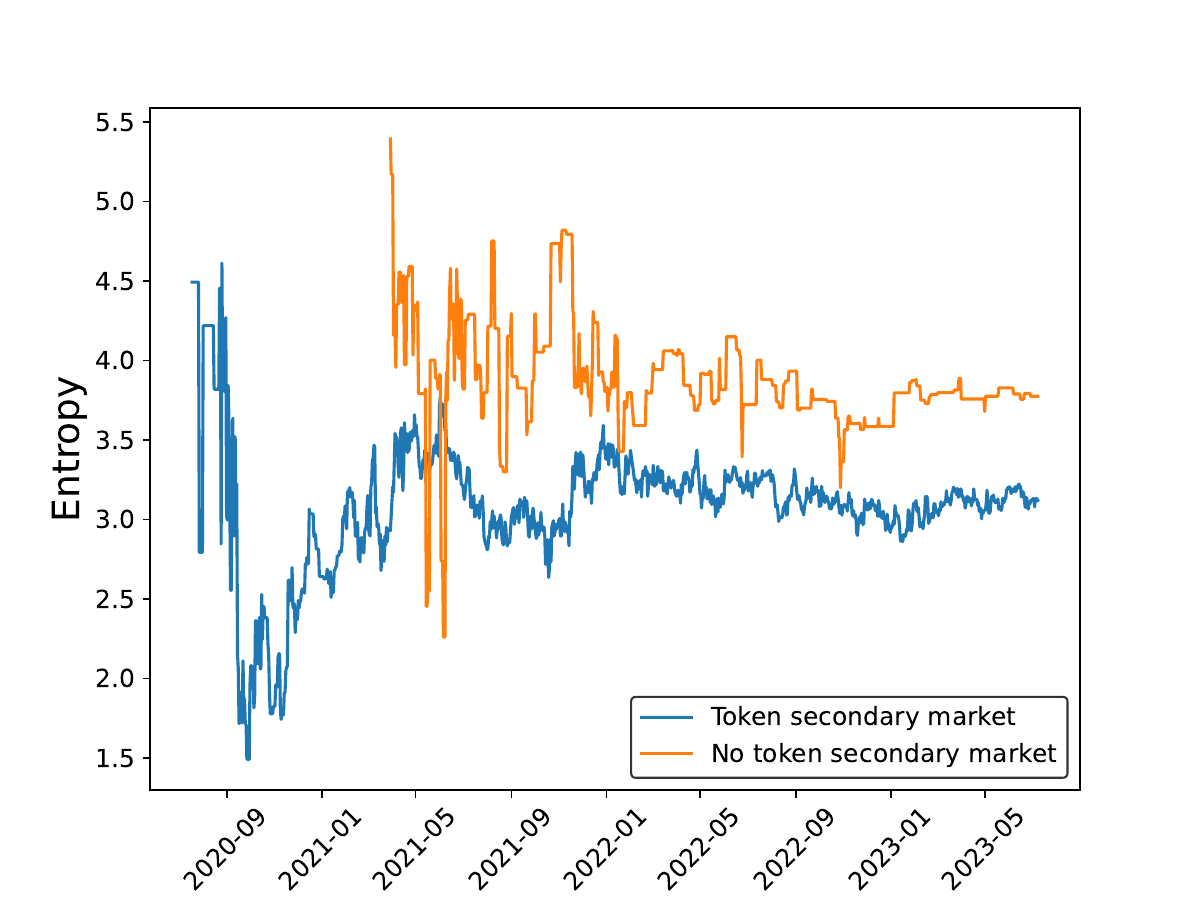}
        \caption{Demonstration with entropy vs DAO with Secondary Market Value}
        \label{fig:login}
    \end{subfigure}%
    ~~
    \begin{subfigure}[t]{0.30\textwidth}
        \centering
        \includegraphics[width=\linewidth]{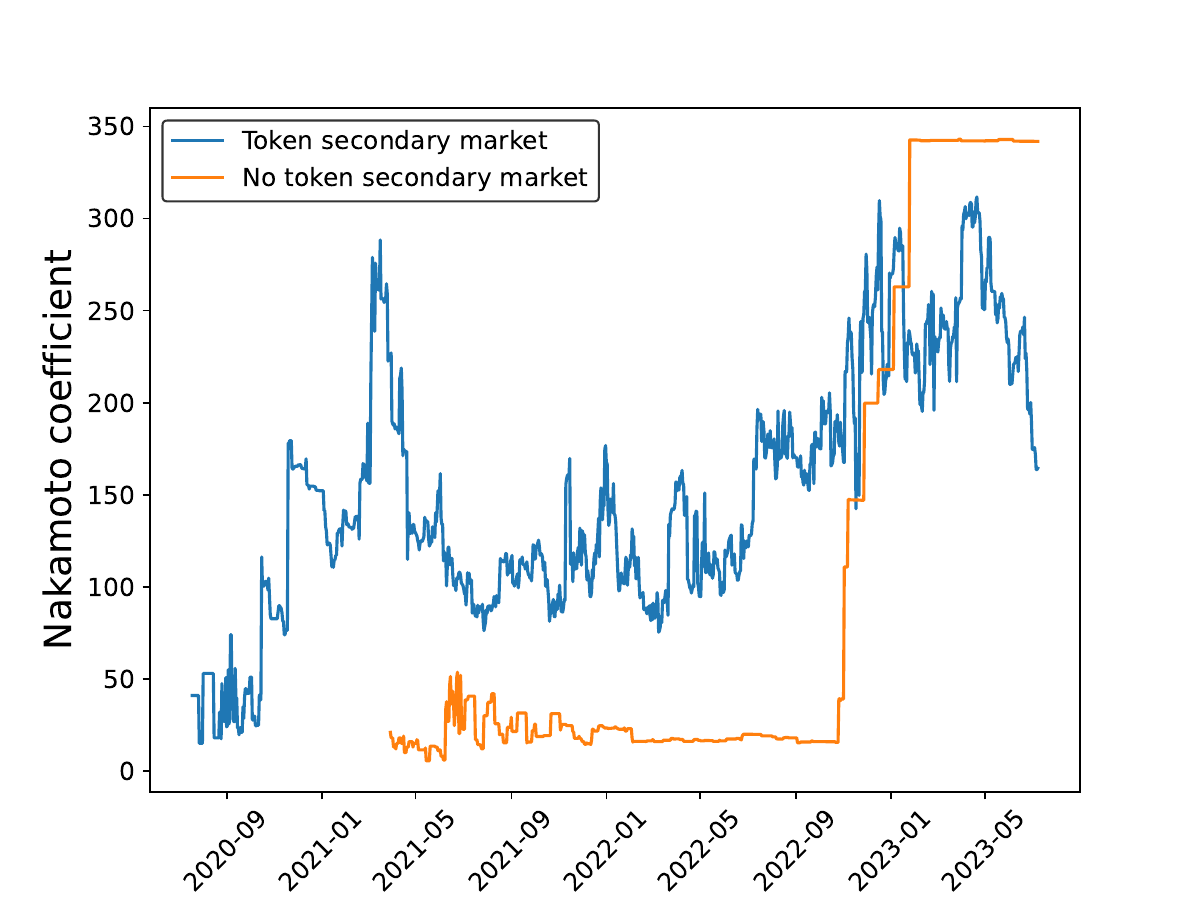}
        \caption{Demonstration with Nakamoto coefficient vs DAO with Secondary Market Value.}
        \label{fig:recovery}
    \end{subfigure}
    \caption{Demonstration of level of decentralization level overtime with Decentralization Metrics}
    \label{fig:dynamic-cluster-secondary}
\end{figure*}
\subsection{Summary of 100 DAOs}
\label{summry-dao}
Out of 100 DAOs analyzed, most used Snapshot for voting. Aavegotchi employed both Snapshot and Aragon smart contracts, while Uniswap used Snapshot for early voting signals and finalized votes on-chain. A few exceptions, such as RnDAO and LaborDAO, did not use formal voting platforms, opting instead for general polling and community discussions on platforms like Discord. Regarding governance tokens, 15\% had no secondary market value, 67\% had market value, and the rest were unspecified. In terms of treasury size, 42\% of DAOs had treasuries between \$0–\$1M, 21\% between \$1M–\$10M, 17\% between \$10M–\$100M, 14\% above \$100M, and the rest were unspecified.
\subsection{Different DAO Properties Impacting Level of Decentralization}
\textbf{Governance Token with the Secondary Market Value Associated with a Lower Level of Decentralization within a DAO.} We run a series of t-tests to statistically compare a decentralization level between two groups: 1) DAOs with a secondary token market and 2) DAOs without a secondary token market. We measure the decentralization level using the three metrics: entropy, Gini, and Nakamoto coefficient. We find that both the average Gini index over time and the current Gini index have a significantly higher value in the first group consisting of DAOs with the secondary token market, indicating a lower degree of decentralization in terms of voting power distribution (average Gini index: $t\approx 5.049,$ $p$-value$<0.001$; current Gini index: $t\approx 3.469,$ $p$-value$<0.001$), in Figure~\ref{fig:secondary}. Similarly, the result of the average Gini with DAO users' participation rate weighted also indicates a small p-value ($p$-value$=0.0002$, $t\approx 3.89$).Therefore, DAOs with a secondary token market tend to have a biased voting power distribution. 

We then present the two groups' historical decentralization levels over time, measured by the three metrics, in Figure~\ref{fig:dynamic-cluster-secondary}. In the graph, a data point presents average values across each group at some time point. Unlike the Gini index, the graphs show entropy between the two groups has been similar. Moreover, the relationship between the Nakamoto values of the two groups is not consistent. Before some point in time, the Nakamoto value in the DAOs without a secondary market was lower than the DAOs with a secondary market. 

We also conduct an OLS regression with multiple variables, including the DAO type. Similar to the t-test, the result shows that whether to have a secondary market is a significant factor in determining their average decentralization level in terms of the Gini index (coeff $= 0.3019$, $p$-value $=0.807$). On the contrary, it is shown that whether to have a secondary market is insignificant to determine their current Gini index when considering other factors such as their DAO type (coeff $= 0.0584$, $p$-value $<0.001$). 
As a result, our analysis shows the existence of their secondary market is significantly associated with their voting power distribution.



\begin{figure*}[t!]
    \centering
    \begin{subfigure}[t]{0.35\textwidth}
        \centering
        \includegraphics[width=\linewidth]{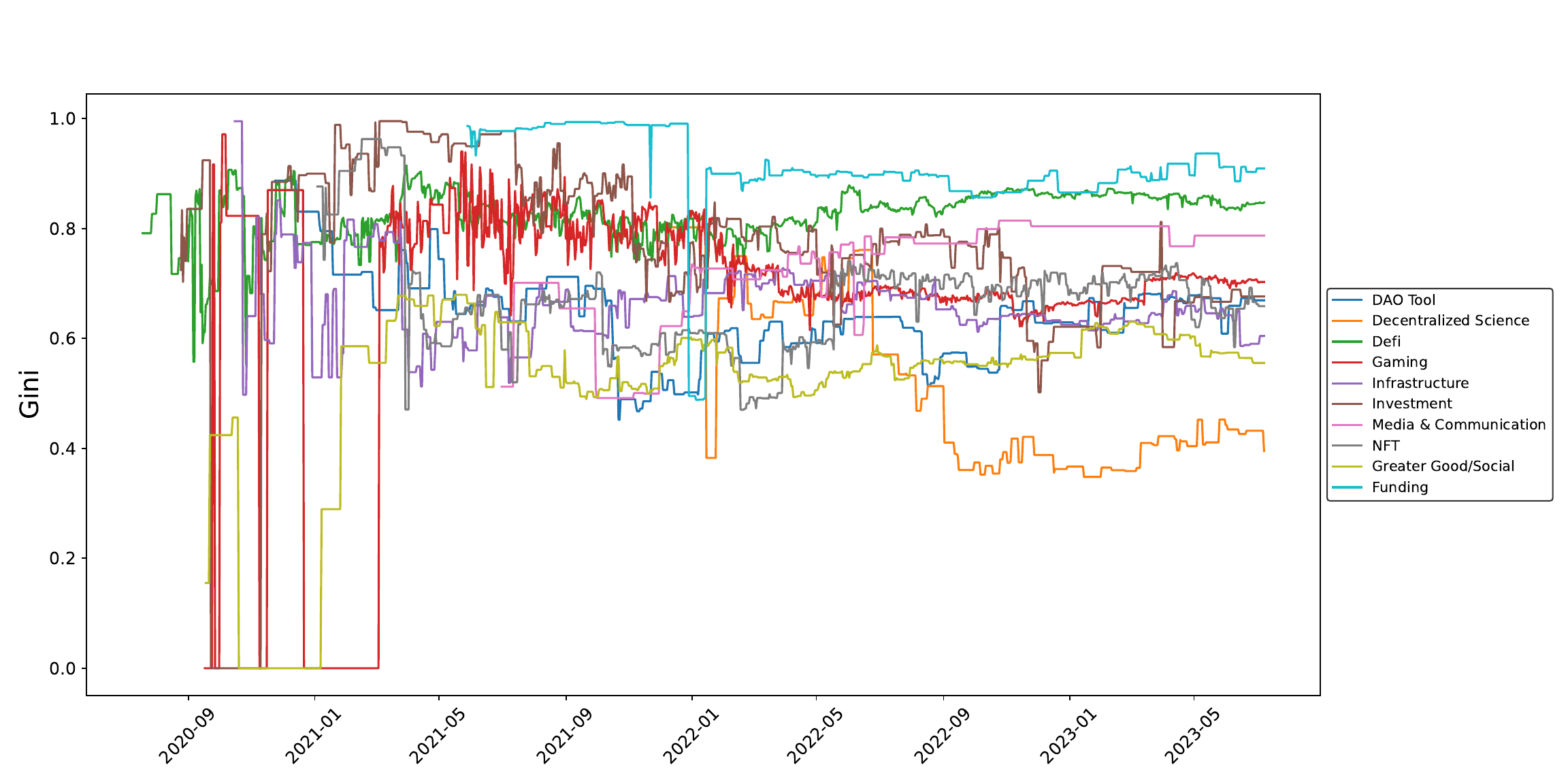}
        \caption{Demonstration with Average Gini Co-efficient vs DAO type}
        \label{fig:registration}
    \end{subfigure}%
    ~~
    \begin{subfigure}[t]{0.35\textwidth}
        \centering
        \includegraphics[width=\linewidth]{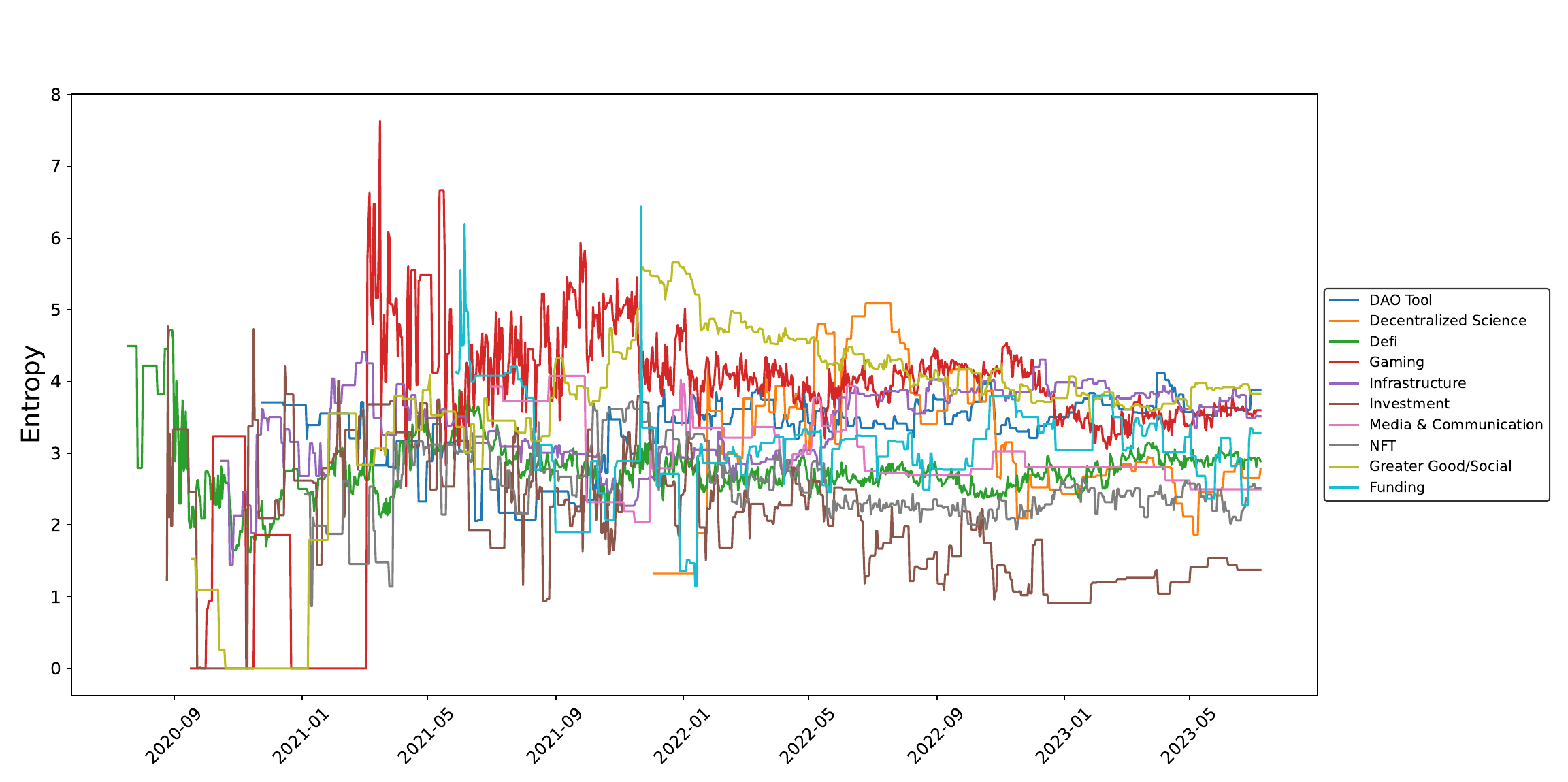}
        \caption{Demonstration with Average antropy Co-efficient vs DAO type}
        \label{fig:login}
    \end{subfigure}%
    ~~
    \begin{subfigure}[t]{0.35\textwidth}
        \centering
        \includegraphics[width=\linewidth]{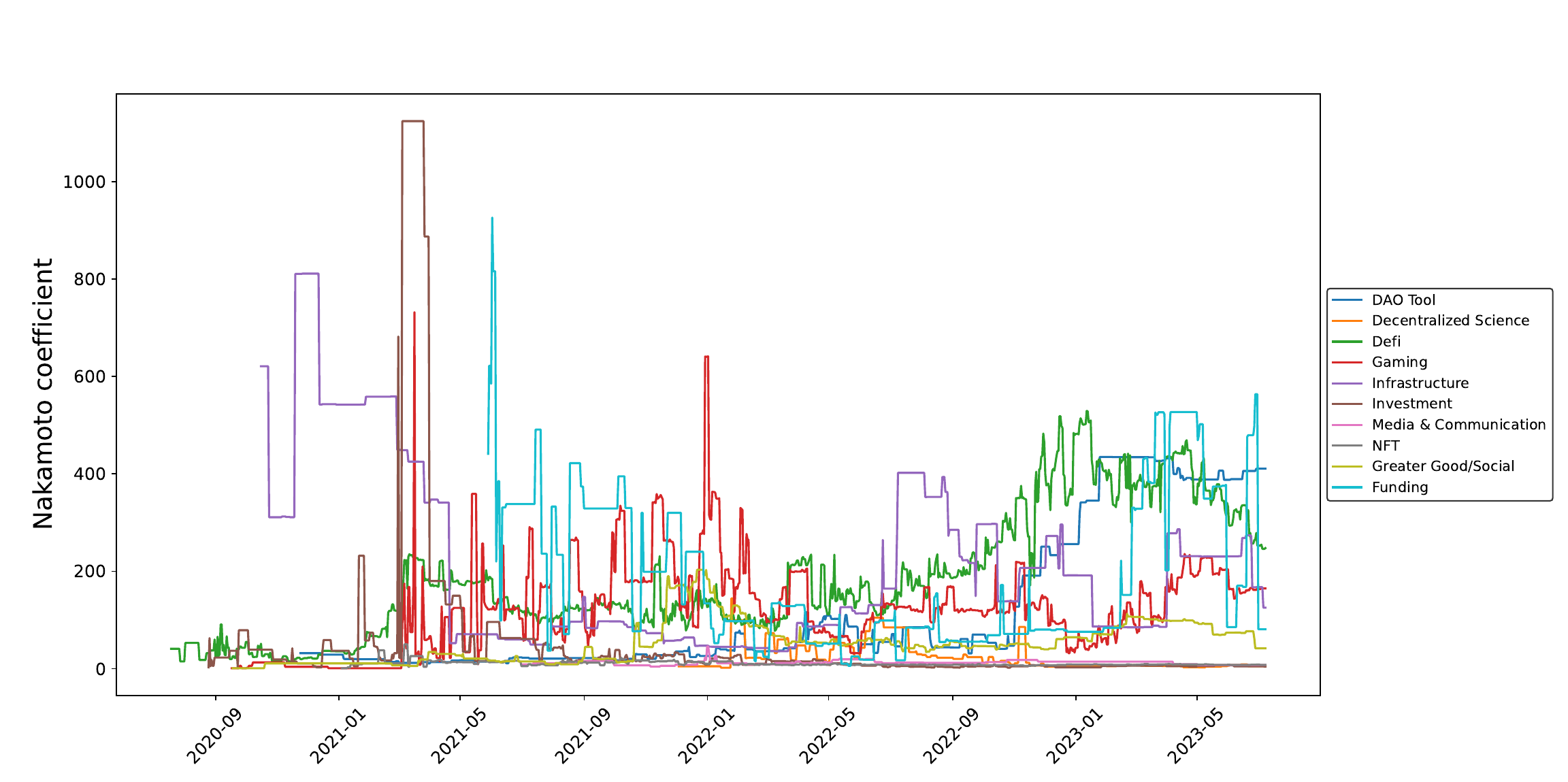}
        \caption{Demonstration with Average nakamotoCo-efficient vs DAO Type}
        \label{fig:recovery}
    \end{subfigure}
    \caption{Demonstration of level of decentralization level overtime with Decentralization Metrics}
    \label{fig:dynamic-cluster-type}
\end{figure*}

\textbf{{Different DAO types are correlated with the level of decentralization.}} We employ ANOVA to statistically compare the levels of decentralization among multiple groups, each representing a distinct category of DAOs. Similar to the analyses above, we use the following three metrics to assess a decentralization level: entropy, the Gini index, and the Nakamoto coefficient. We observed differences in decentralization levels across these various categories. Figure~\ref{fig:dynamic-cluster-type} presents a historical decentralization level for various DAO types. It shows that research, data analysis, social/greater good, and sports-oriented DAOs exhibit lower Gini values, indicating higher decentralization. In contrast, infrastructure, funding, product, and investment-oriented DAOs display higher Gini values, suggesting a lower degree of decentralization. Figure~\ref{fig:average-decen-type} in the Appendix also presents the same results. Relatedly, our analysis of the ANOVA test yields a Gini coefficient with a p-value of $0.05$. 

Moreover, when we run an OLS analysis with multiple variables, the social/greater good DAO type is a significant factor in determining the Gini index (coeff$=1.1545,$ $p$-value $=0.028$). On the other hand, the DeFi and investment DAO types are significant factors regarding the Gini index (DeFi: coeff$=0.1121,$ $p$-value $=0.003$; Investment: coeff$=0.1273,$ $p$-value $=0.070$). As a result, we can conclude that the category of DAOs has an impact on decentralization levels, suggesting potential differences in entropy and Gini coefficient across different types of DAOs. It is important to note that various DAOs may belong to multiple categories, such as Radicle, API3, Osmosis, and Unlock, which may fall into both the DAO tools and Infrastructure categories, adding complexity to the analysis.

\vspace{-2mm}
\subsection{Evaluation of Decentralization Levels with Voting Power}

\textbf{Pearson Correlation.} We conduct a correlation analysis between the voting participation rate of a user and its voting power. If it shows a significant positive correlation, it can be a negative sign of a decentralization level cause more powerful voters tend to participate in voting more actively, underrepresenting minor voters more. In the analysis, we calculate a user’s average voting power and the number of proposals that the user participated in voting. Then calculate a Pearson correlation coefficient. Table~\ref{tab:voting_metric-1} summarizes a Pearson correlation coefficient and p-value of 100 DAOs. While the results were mixed, DAOs like 1inch, save, arbitrum, cakevote, club, fei, freerossdap, gmx, klima,  gnosis, humanDAO, kcc, olympus, pie, sushiswap, etc - showed a significant positive correlation. This means that in those DAOs more powerful voters tend to participate in voting more actively, leading to a poor level of decentralization. Consequently, it appears that in many DAOs, wealthy voters are more likely to participate, exacerbating poor decentralization.
LeagueDAO, partyDAO, peopleDAO, and treasure gaming, on the other hand, displayed a significantly strong negative correlation where participation from whale voters is not skewed. 

\begin{figure*}[ht]
    \centering
    \begin{subfigure}[t]{0.23\textwidth}
        \centering
        \includegraphics[width=\linewidth]{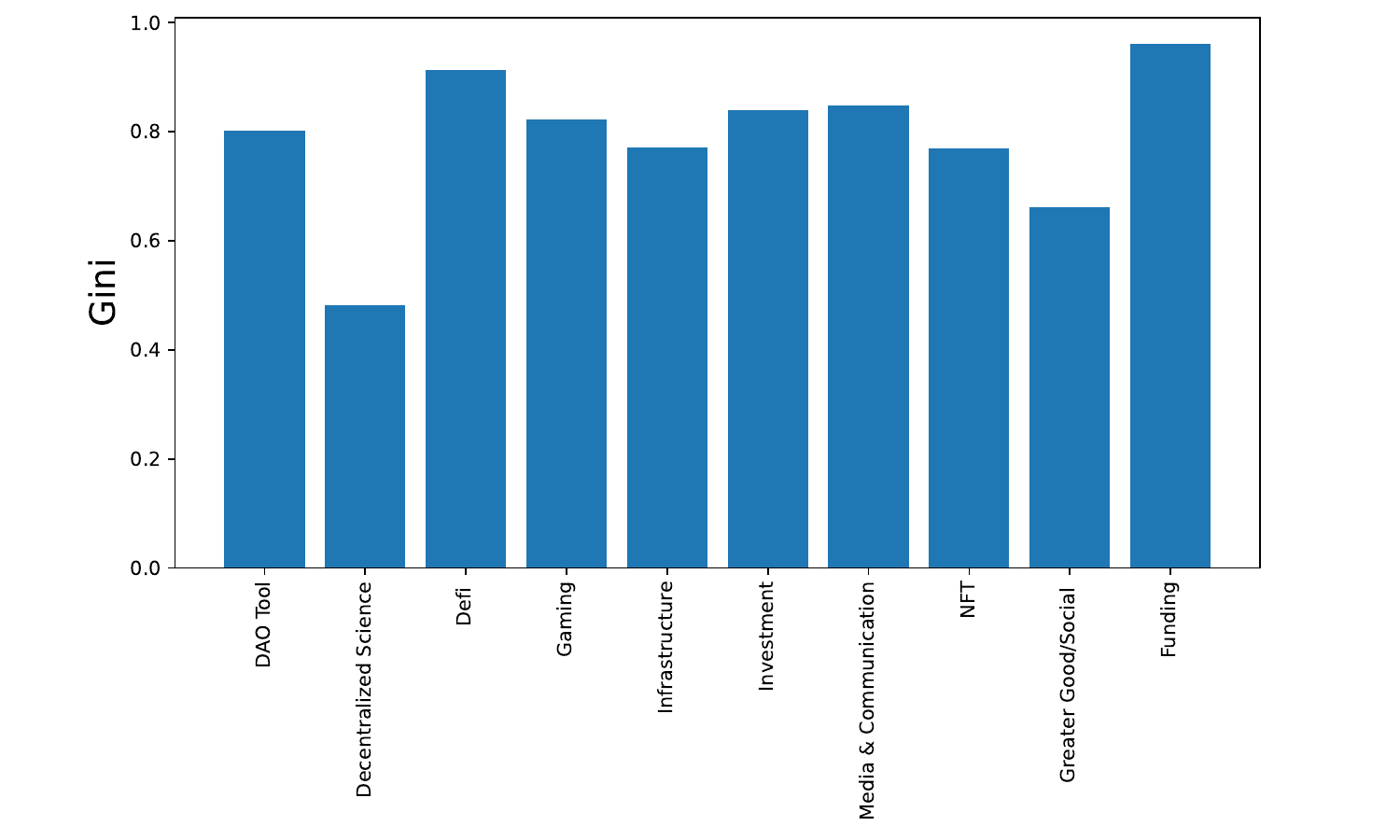}
        \caption{average gini vs type}
        \label{fig:registration}
    \end{subfigure}%
    ~~
    \begin{subfigure}[t]{0.23\textwidth}
        \centering
        \includegraphics[width=\linewidth]{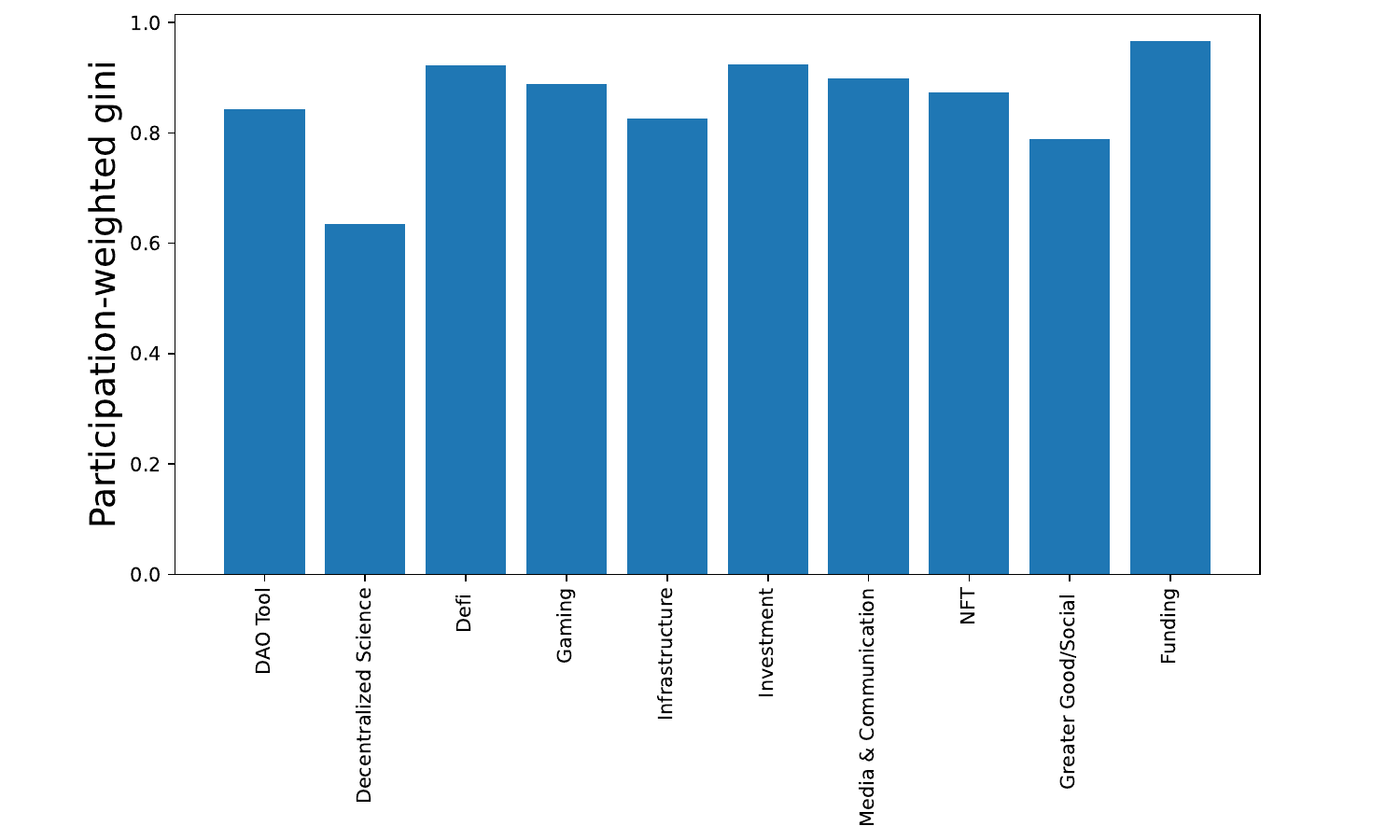}
        \caption{average participation gini vs type}
        \label{fig:login}
    \end{subfigure}%
    ~~
    \begin{subfigure}[t]{0.23\textwidth}
        \centering
        \includegraphics[width=\linewidth]{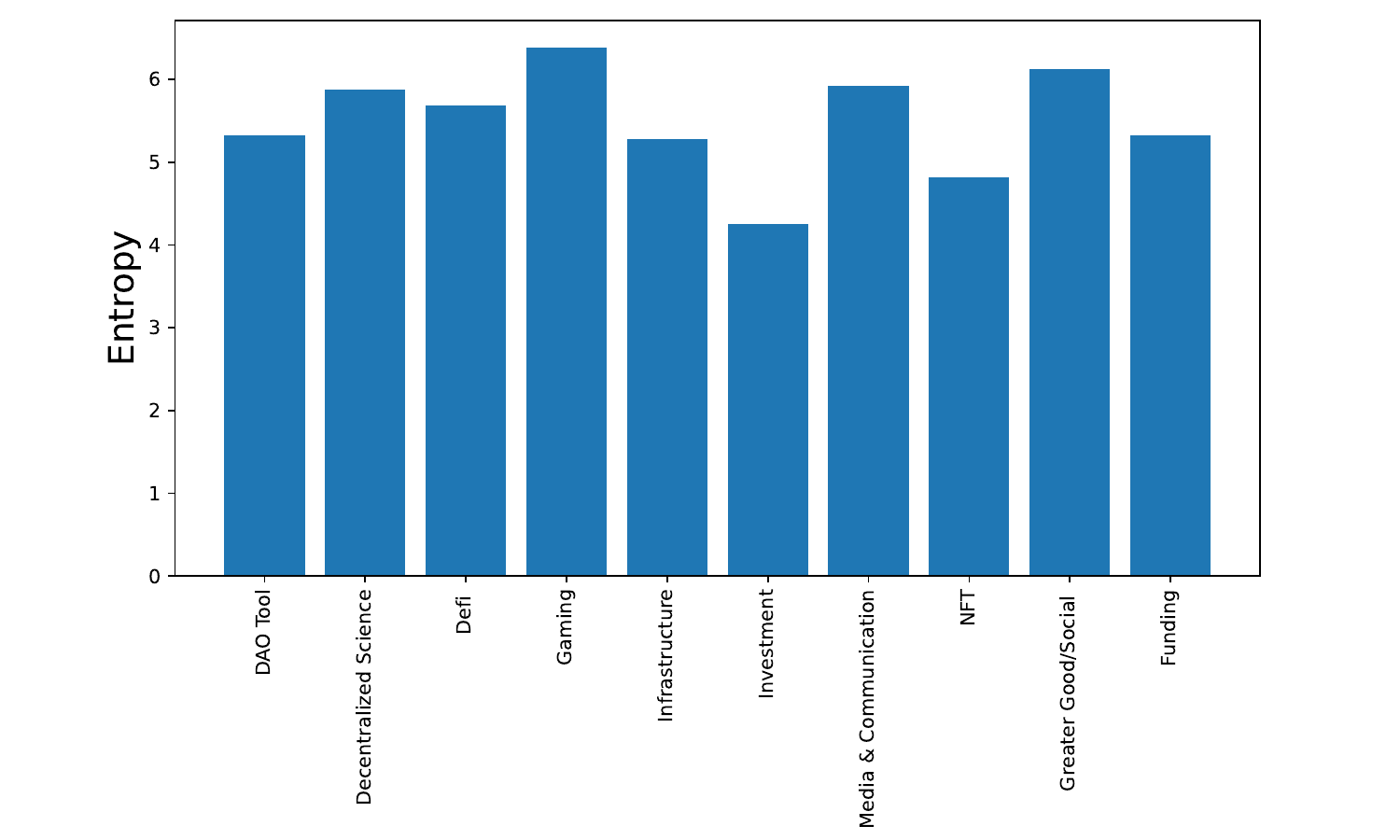}
        \caption{average entropy vs type}
        \label{fig:recovery}
    \end{subfigure}
    ~~
    \begin{subfigure}[t]{0.23\textwidth}
        \centering
        \includegraphics[width=\linewidth]{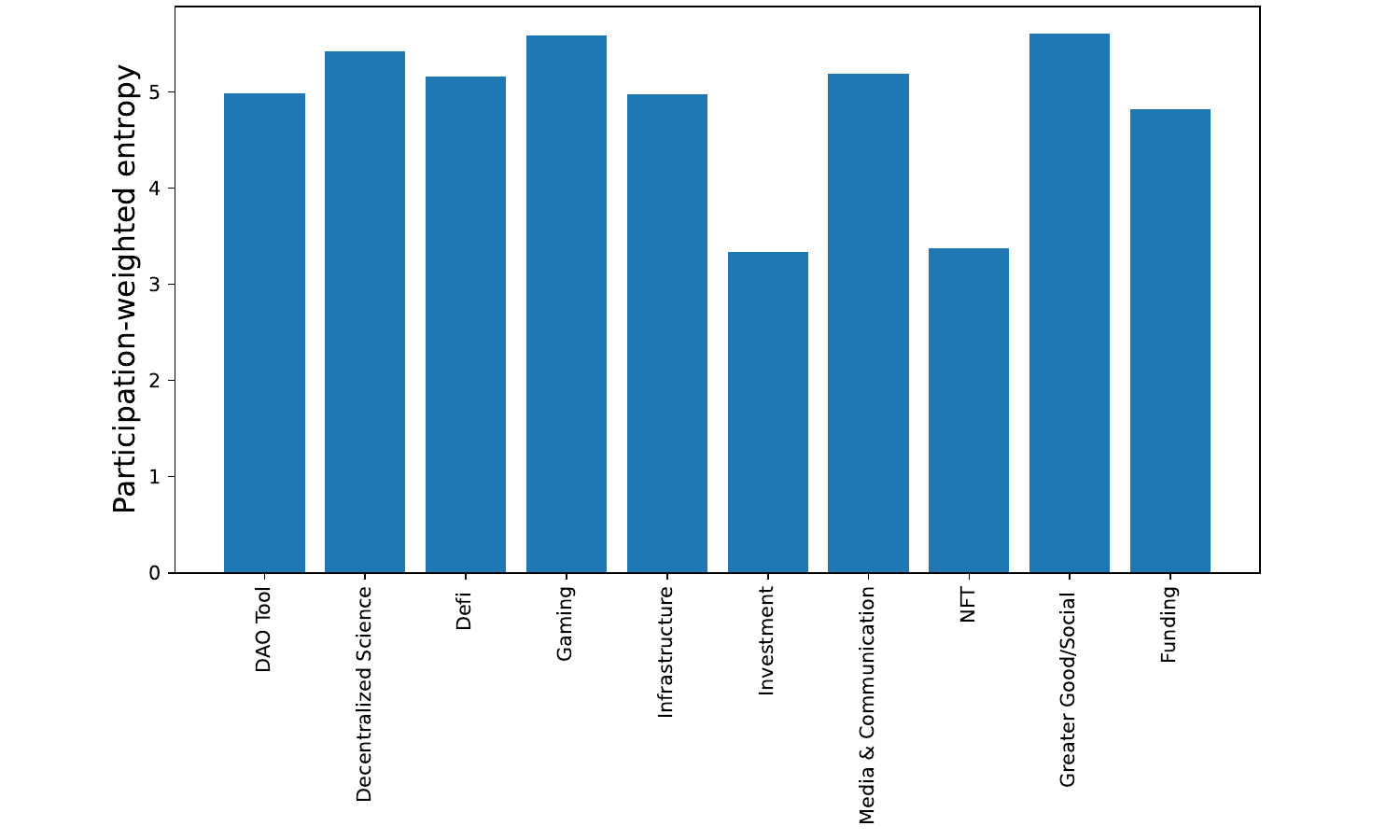}
        \caption{average participation entropy vs type}
        \label{fig:recovery}
    \end{subfigure}
    \caption{Average decentralization level (DAO Type) with Decentralization Metrics}
    \label{fig:average-decen-type}
\end{figure*}
\textbf{Decentralization Level (Gini, Entropy, Nakamoto).} Next, we analyze a decentralization level of voting mechanisms using three metrics: entropy, Gini, and Nakamoto. Before applying the three metrics, we first define the “recognized power” of a user as how much voting power a user has actually exercised. For example, even if a user has a lot of tokens that can be interpreted as voting power if the user rarely participates in voting in general, some may think they are not influential entities in a voting mechanism. Given the rationale, we calculate the “recognized power” as $V*P$, where V is a user’s voting power (or token amounts) and P is the user’s participation ratio. 

We apply entropy and Gini to both voting power and the recognized power of users. On the other hand, Nakamoto is applied to only voting power because Nakamoto is to estimate the minimum number of users who can threaten a voting mechanism under the assumption that the users are corrupted. Entropy is calculated as $-\sum_i p_i log2(p_i)$, where $p_i$ is a voting power (or recognized power) of user i / the total voting powers (or recognized powers) of all users. Gini is calculated as $\frac{\sum_i \sum_j |p_i - p_j|}{2\sum_i \sum_j p_j}$. For Nakamoto, $\min_n (\sum_{i=1}^{n} q_i) >= Q$, where $q_i$ is user i’s voting power and Q is a quorum threshold, the minimum amount of voting power for a proposal to pass.

From Table~\ref{tab:voting_metric-1}, we can see many DAOs suffer from poor decentralization. In particular, 1inch, save, balancer, dydx, fei, gmx, gnosis, julswap, kogecoin. klime, lido, etc showed significantly poor decentralization, where their Gini coefficients were above 0.9 due to an extremely large power inequality or (and) the Entropy values were low due to a low number of participating voters. On the other hand,  tosinshada, metricsdao, lexdao,  olympus, goldhunt, golflinks, giaidao, graph protocol, daocity, friends with benefit showed the highest decentralization level among our target DAOs.

\begin{figure*}[t!]
    \centering
    \begin{subfigure}[t]{0.25\textwidth}
        \centering
        \includegraphics[width=\linewidth]{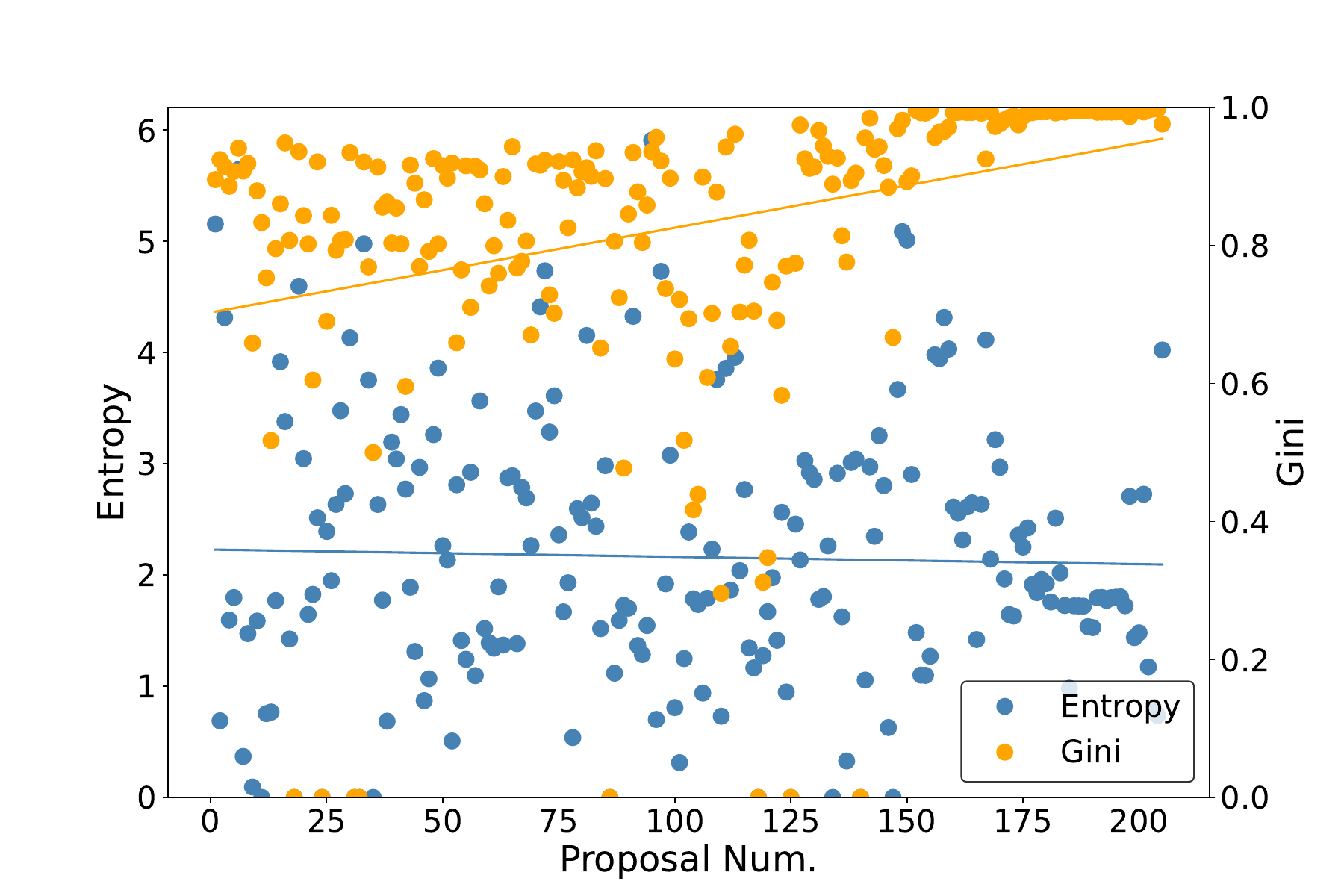}
        \caption{Sushiswap}
        \label{fig:registration}
    \end{subfigure}%
    ~~
    \begin{subfigure}[t]{0.25\textwidth}
        \centering
        \includegraphics[width=\linewidth]{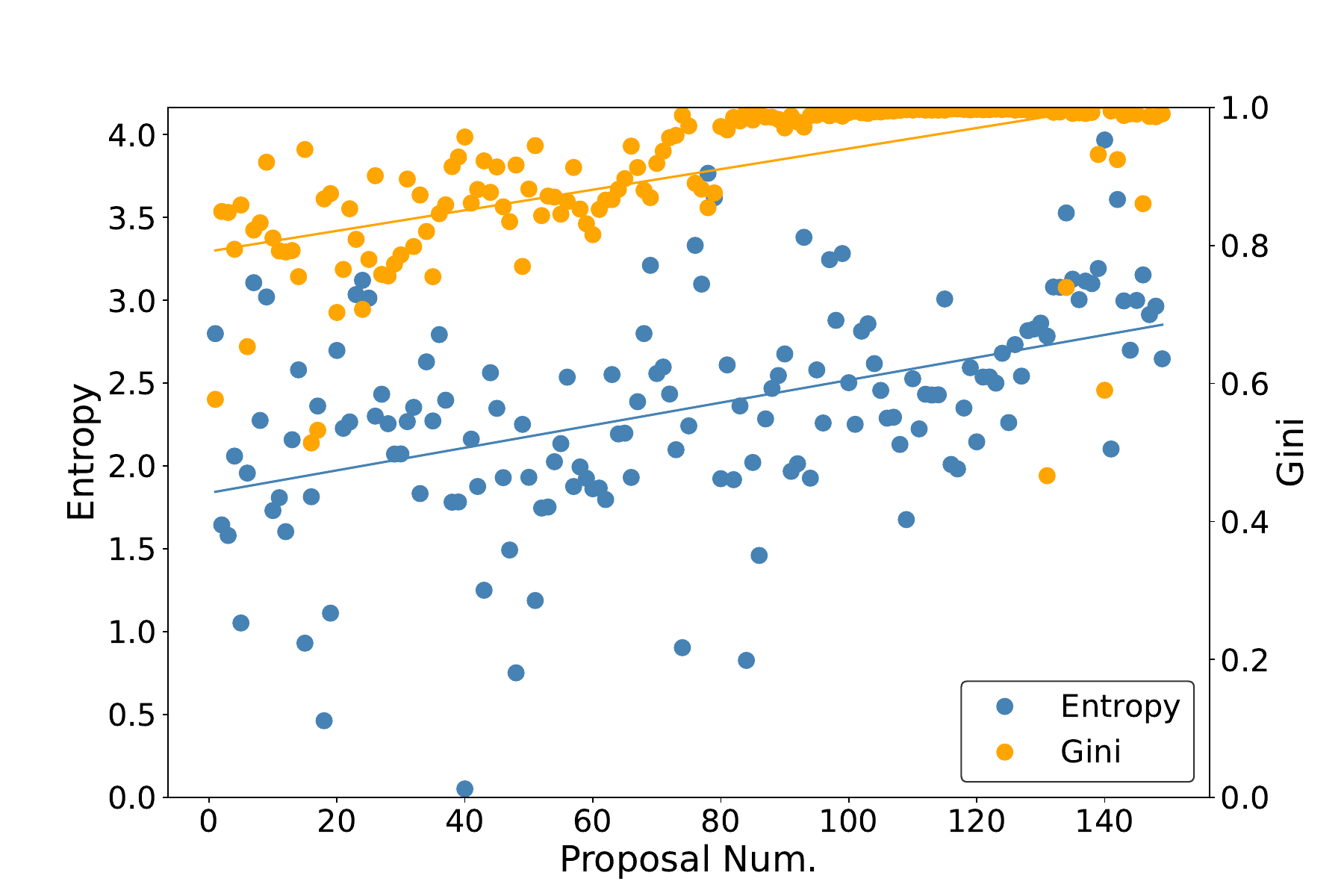}
        \caption{Lido}
        \label{fig:login}
    \end{subfigure}%
    ~~
    \begin{subfigure}[t]{0.25\textwidth}
        \centering
        \includegraphics[width=\linewidth]{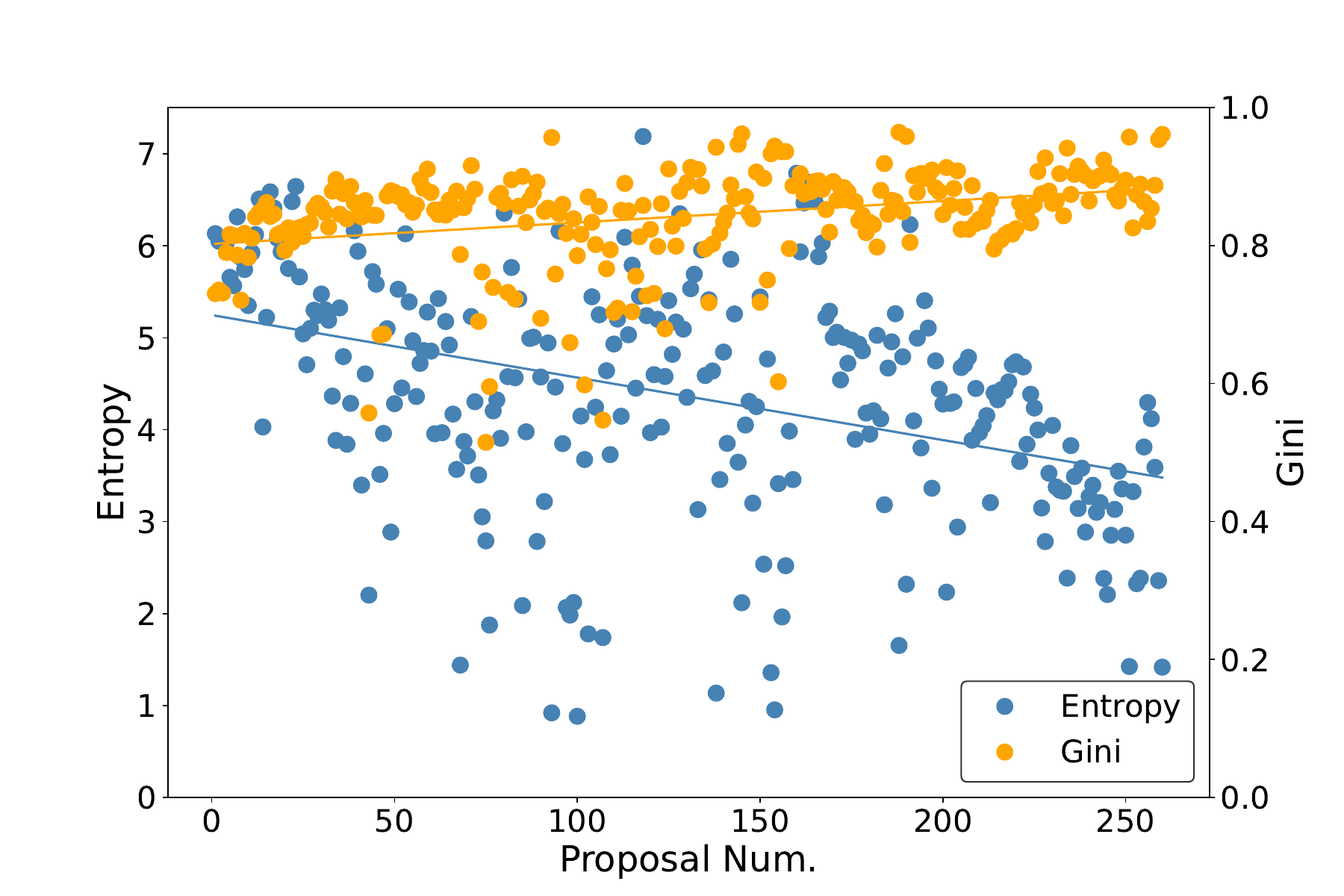}
        \caption{Olympus DAO}
        \label{fig:recovery}
    \end{subfigure}
    ~~
    \begin{subfigure}[t]{0.25\textwidth}
        \centering
        \includegraphics[width=\linewidth]{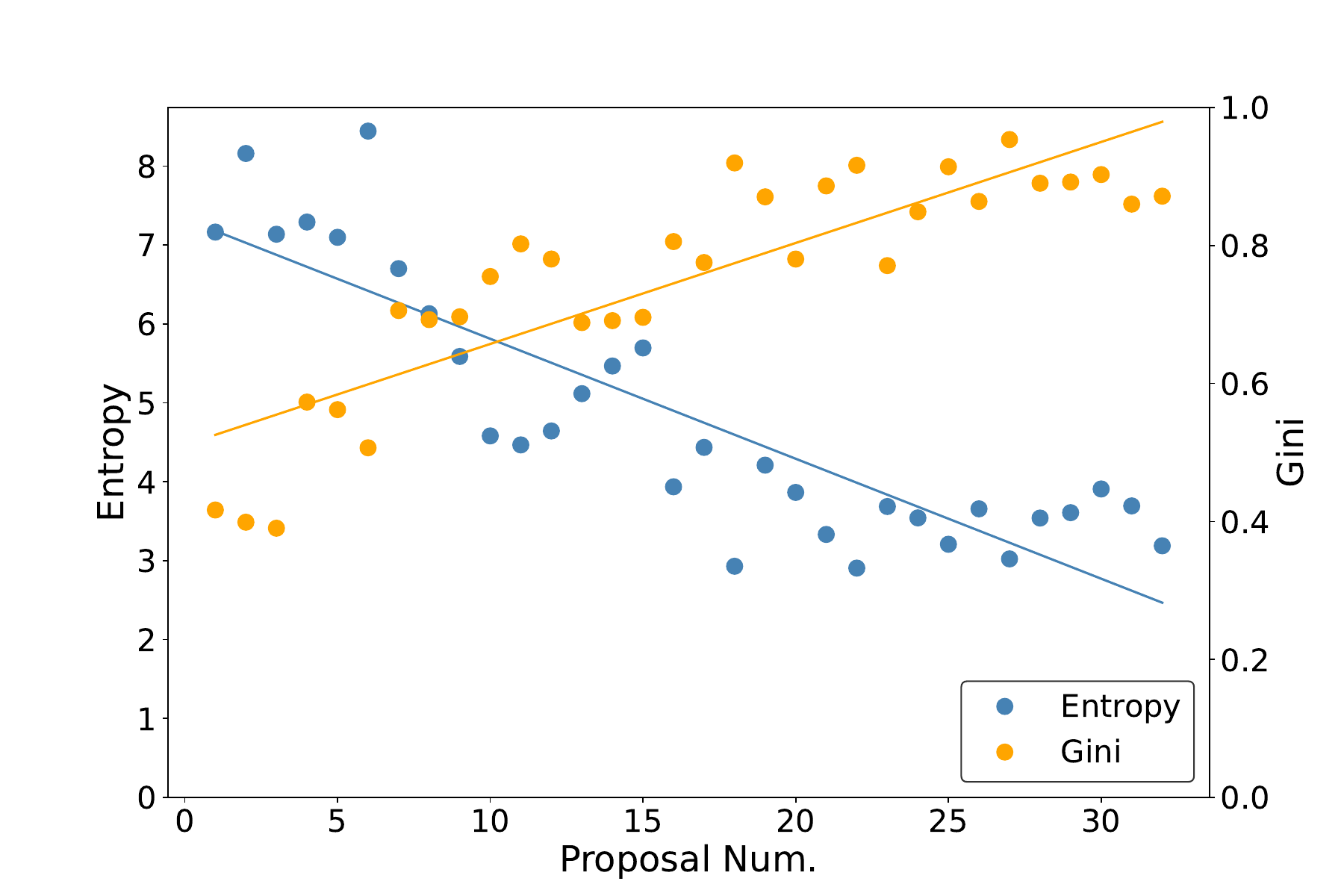}
        \caption{Ribbon}
        \label{fig:recovery}
    \end{subfigure}
 
    \caption{Decentralization Level in each DAO, demonstrating level of decentralization increase and decrease over time for some example DeFi DAO and Social DAOs}
    \label{fig:social-defi}
\end{figure*}

\begin{figure*}[t!]
    \centering
    \begin{subfigure}[t]{0.25\textwidth}
        \centering
        \includegraphics[width=\linewidth]{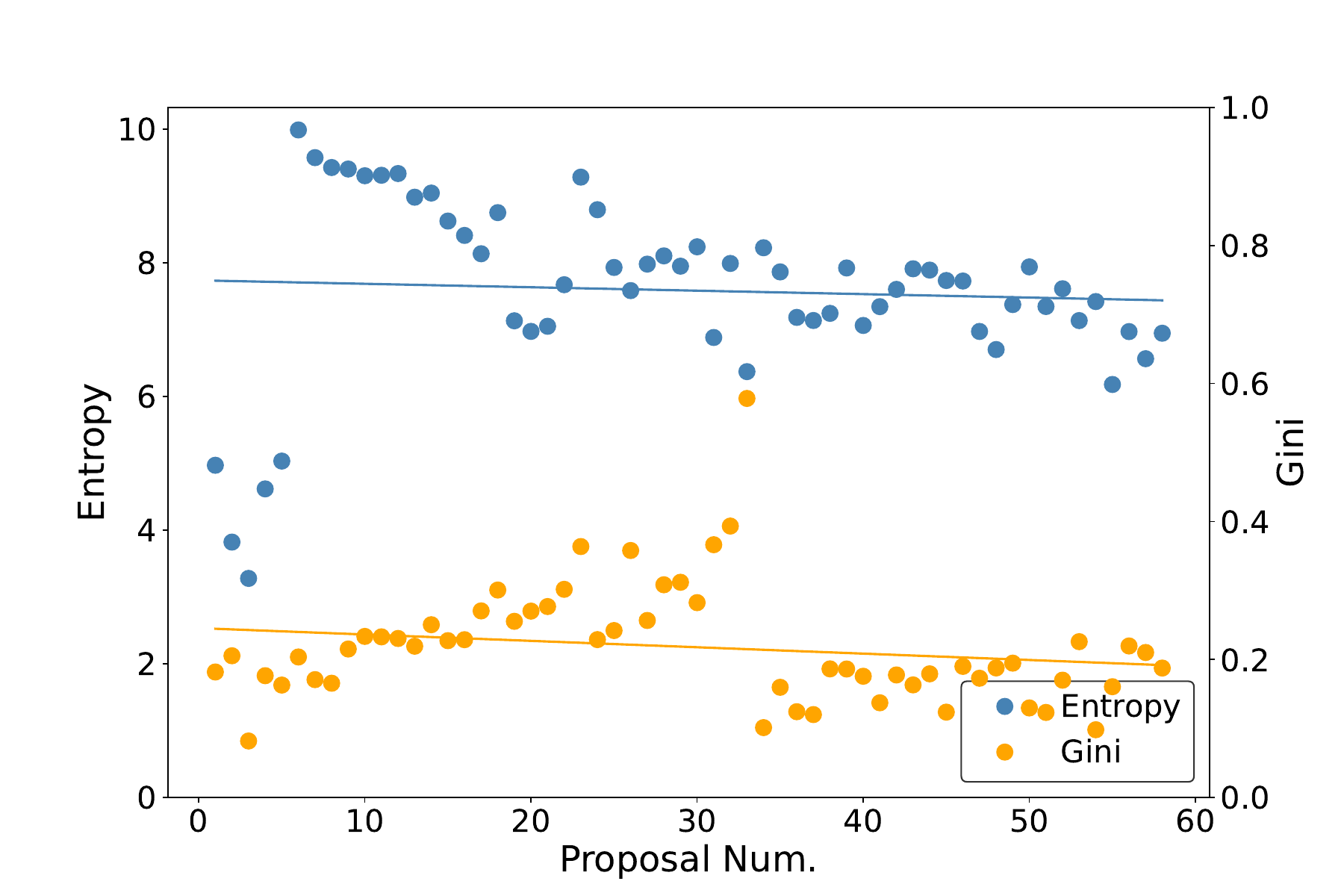}
        \caption{CityDAO}
        \label{fig:recovery}
    \end{subfigure}
     ~~
    \begin{subfigure}[t]{0.25\textwidth}
        \centering
        \includegraphics[width=\linewidth]{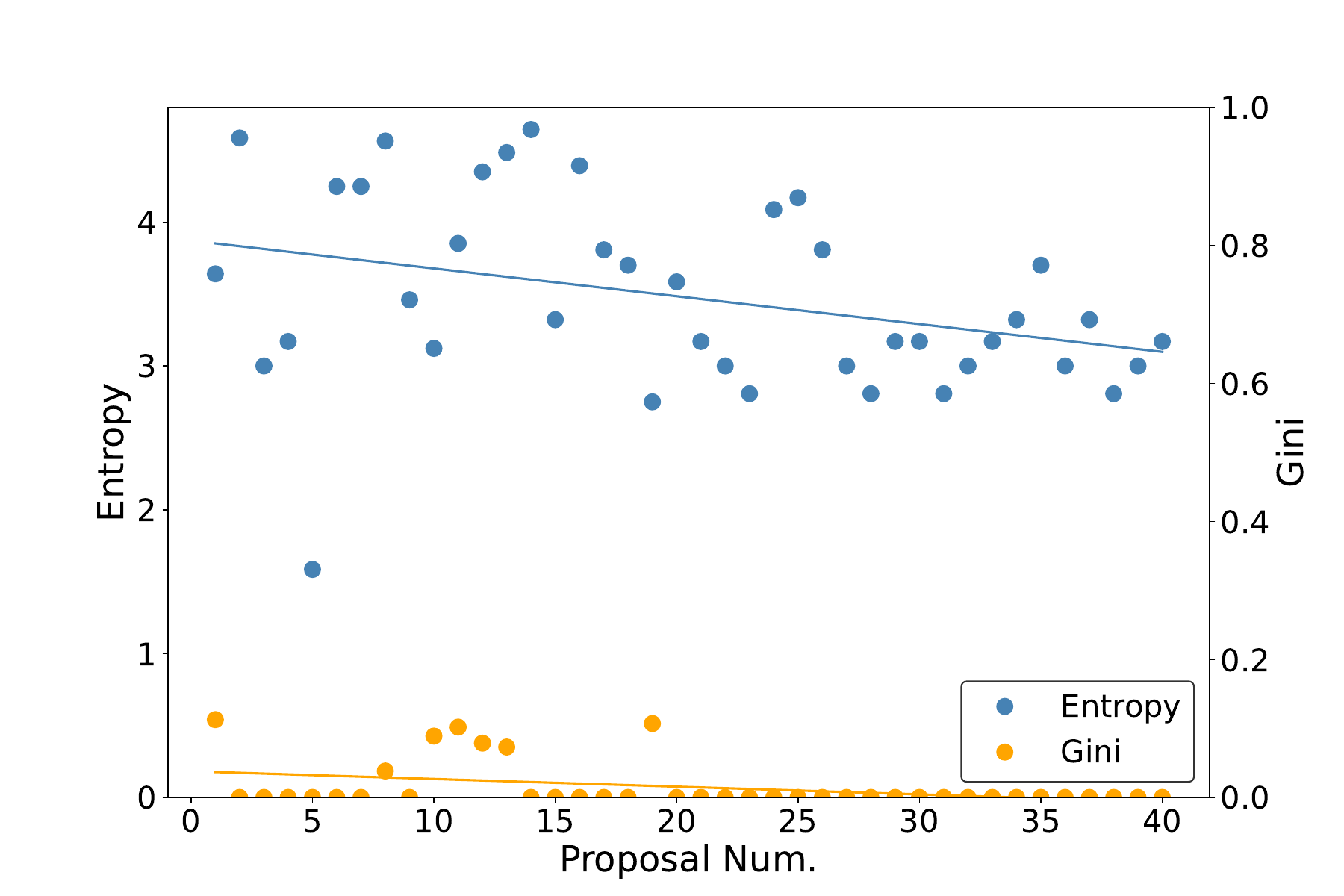}
        \caption{LexDAO}
        \label{fig:recovery}
    \end{subfigure}
        ~~
    \begin{subfigure}[t]{0.25\textwidth}
        \centering
        \includegraphics[width=\linewidth]{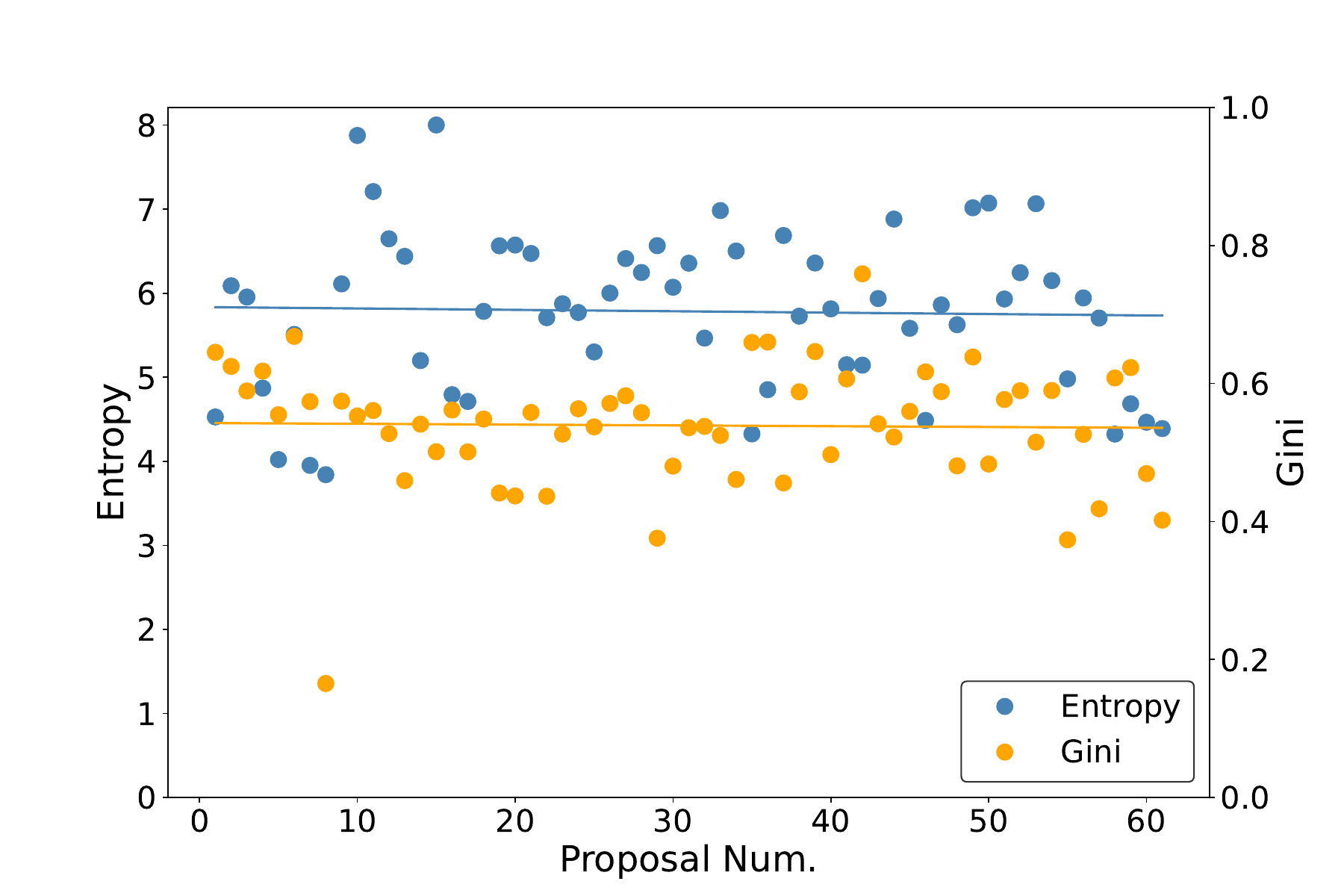}
        \caption{Friends with Benefits}
        \label{fig:recovery}
    \end{subfigure}
        ~~
    \begin{subfigure}[t]{0.25\textwidth}
        \centering
        \includegraphics[width=\linewidth]{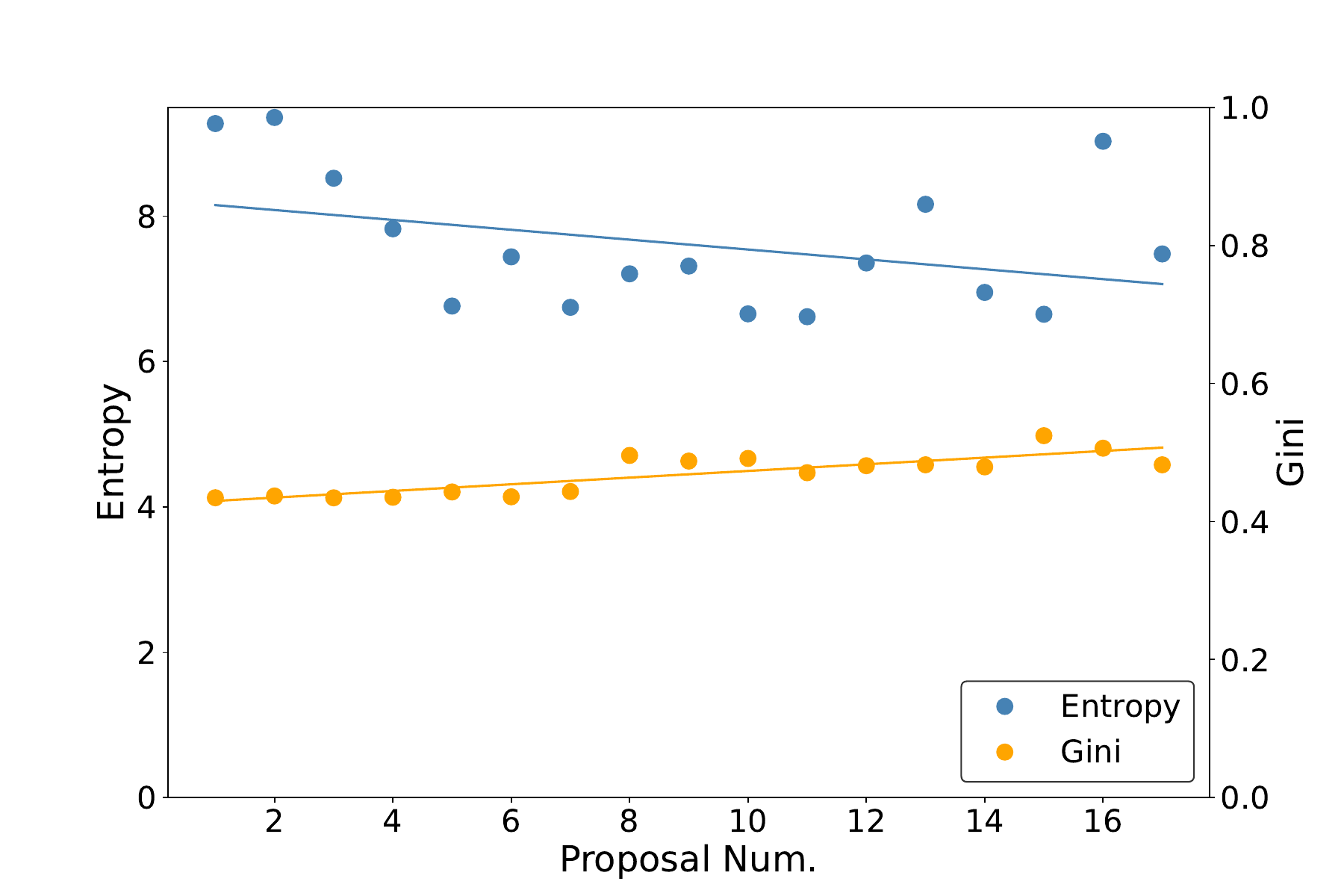}
        \caption{LinksDAO}
        \label{fig:recovery}
    \end{subfigure}
    \caption{Decentralization Level in each DAO, demonstrating level of decentralization increase and decrease over time for some example DeFi DAO and Social DAOs}
    \label{fig:social-defi2}
\end{figure*}

\begin{figure*}[t!]
    \centering
    \begin{subfigure}[t]{0.25\textwidth}
        \centering
        \includegraphics[width=\linewidth]{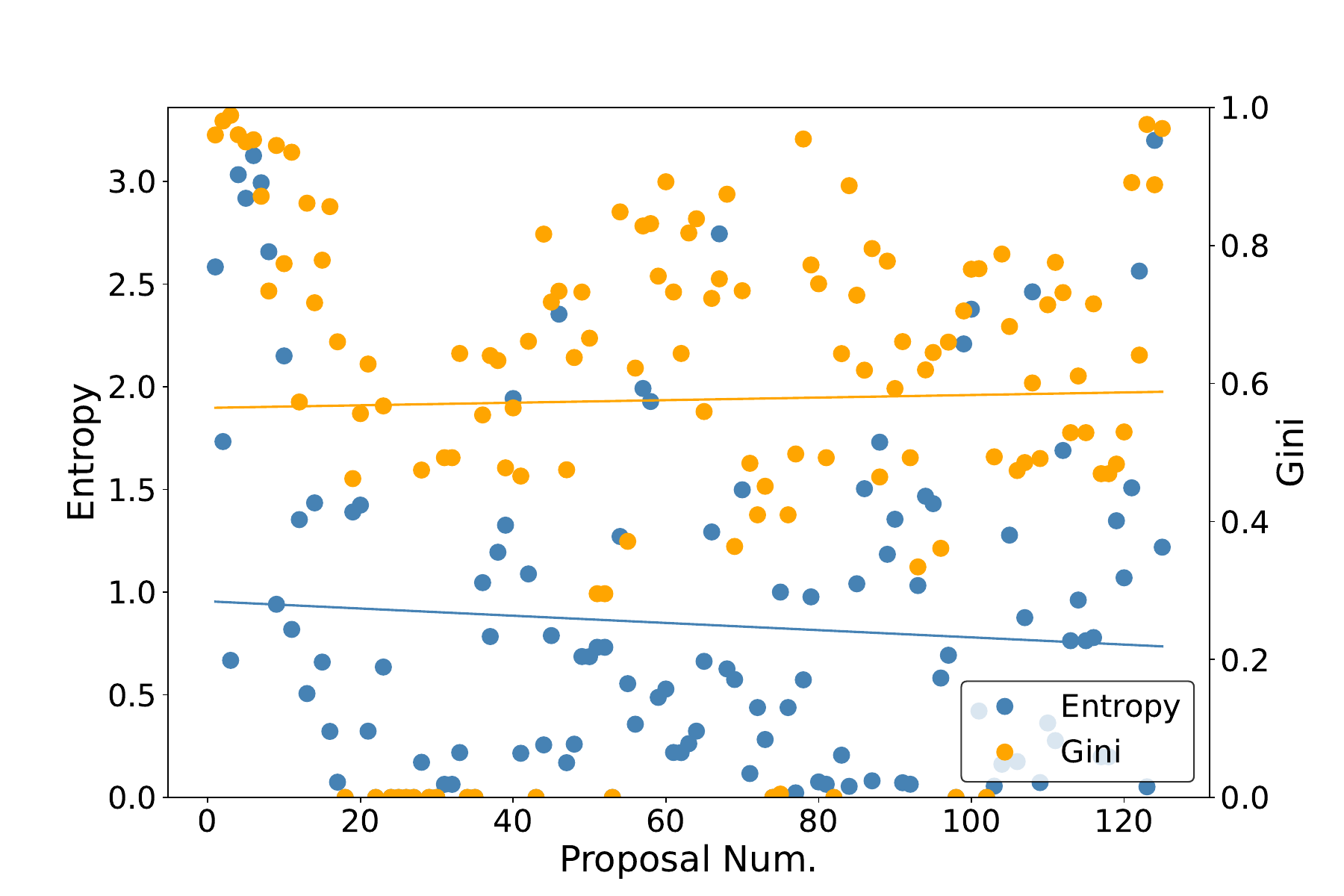}
        \caption{curvedao Defi Type}
        \label{fig:registration}
    \end{subfigure}%
    ~~
    \begin{subfigure}[t]{0.25\textwidth}
        \centering
        \includegraphics[width=\linewidth]{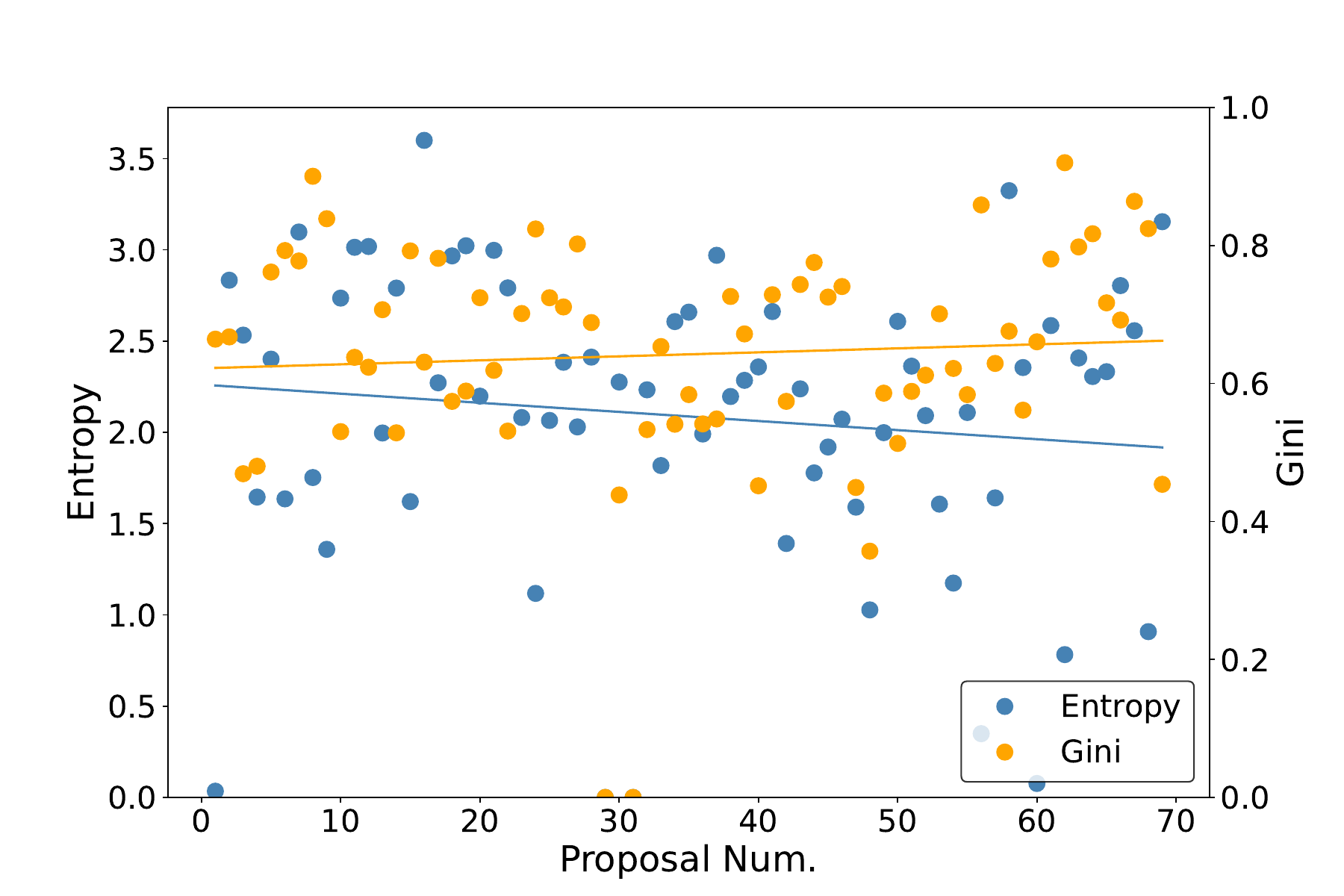}
        \caption{Trufigov Defi Type}
        \label{fig:login}
    \end{subfigure}%
    ~~
    \begin{subfigure}[t]{0.25\textwidth}
        \centering
        \includegraphics[width=\linewidth]{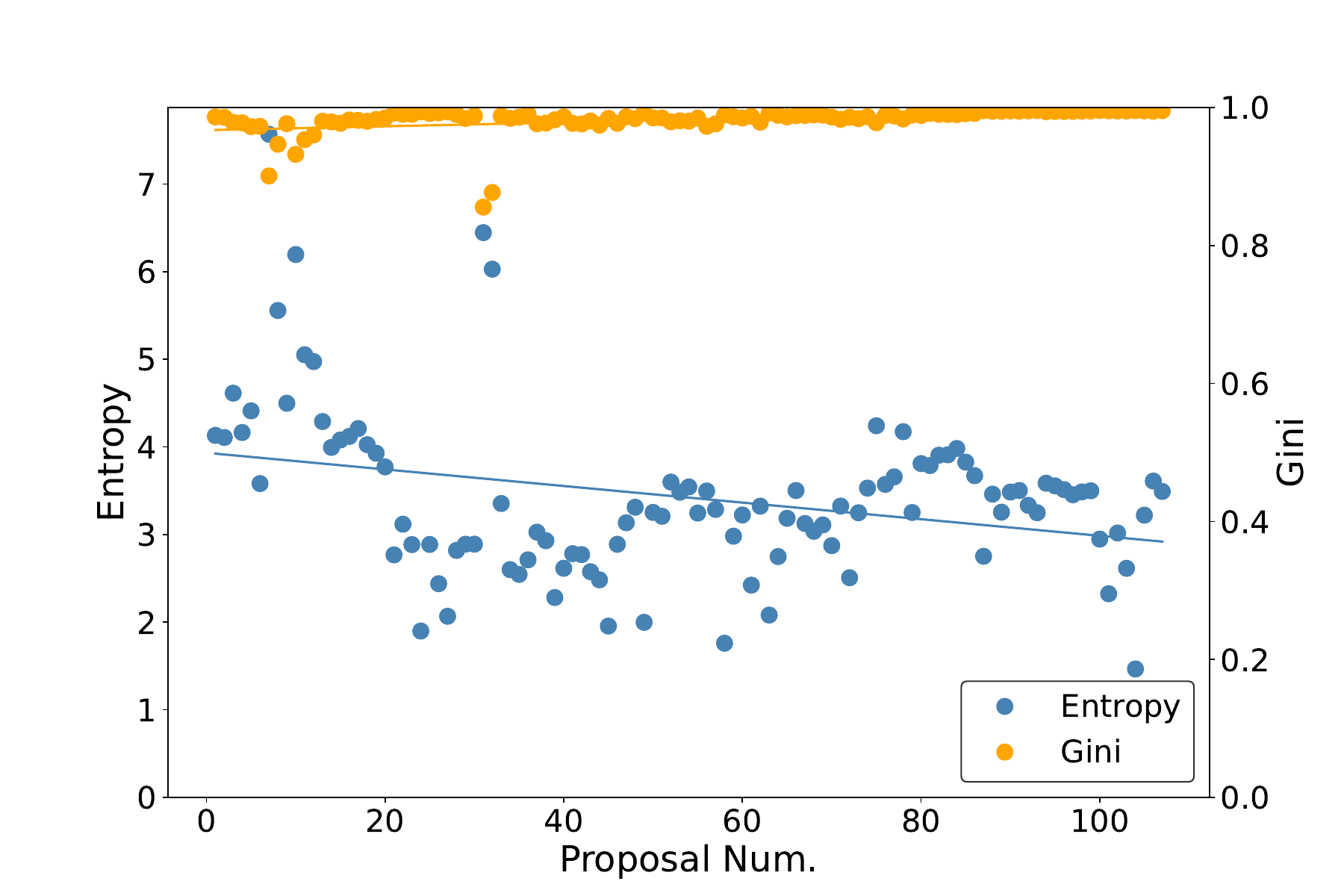}
        \caption{Gitcoin DAO research or funding type}
        \label{fig:recovery}
    \end{subfigure}
    ~~
    \begin{subfigure}[t]{0.25\textwidth}
        \centering
        \includegraphics[width=\linewidth]{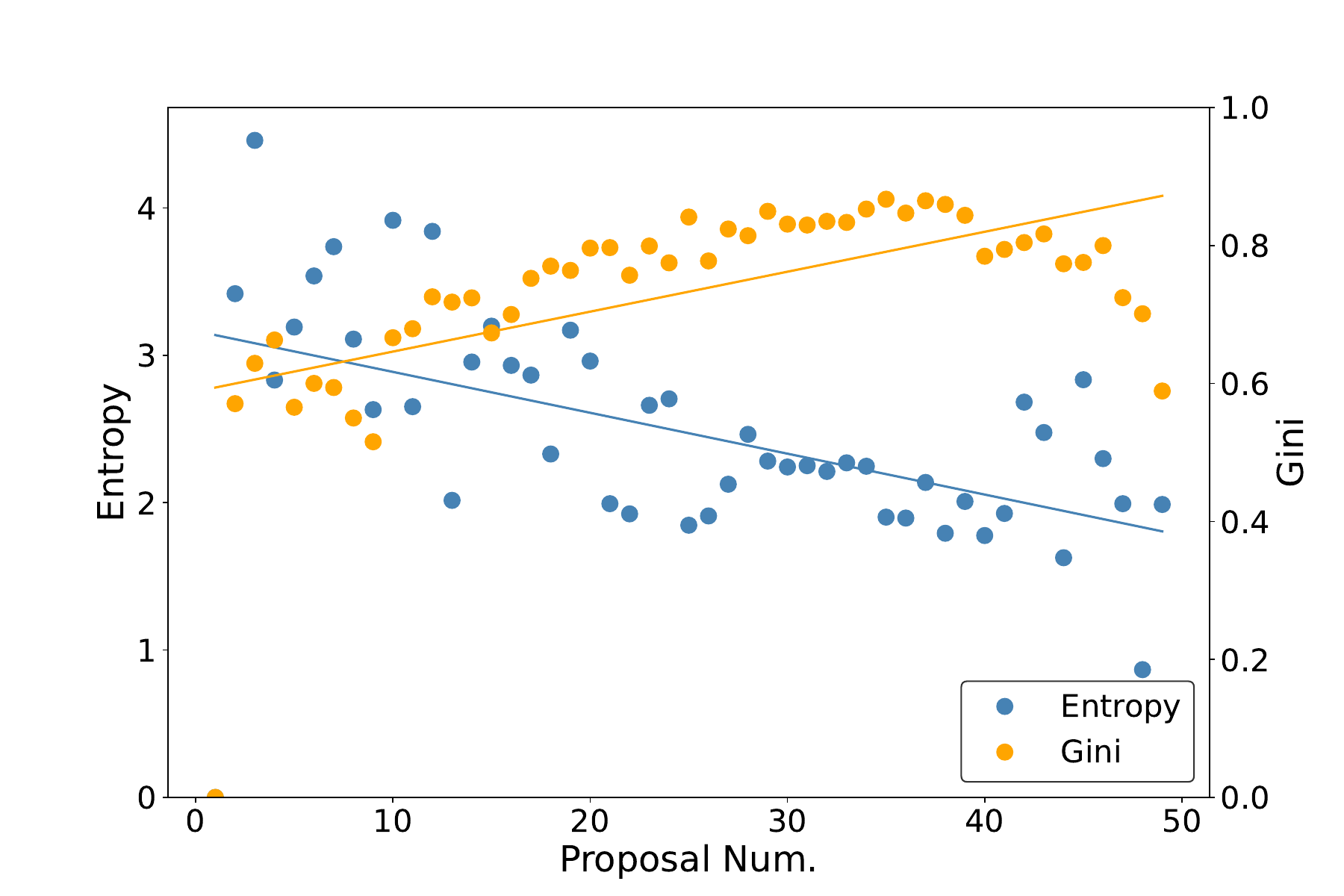}
        \caption{Canbin DAO Social dao type}
        \label{fig:recovery}
    \end{subfigure}
 
    \caption{Some exception of decentralization level where there is an increase/decrease of decentralization regardless of the DAO type}
    \label{fig:social-defi3}
\end{figure*}

\textbf{Decentralization Dynamics Overtime.} 
We also examine the evolution of decentralization dynamics over time. In the Appendix, Table~\ref{tab:voting_metric-1} provides a summary of linear regression results for 100 DAOs, highlighting significantly negative and positive trends in decentralization with orange and blue, respectively. Overall, a notable trend emerges, with a majority of DAOs in the Defi, Investment, and Infrastructure categories showing a consistently negative decentralization trend—indicating a decrease in decentralization levels over time. Figure~\ref{fig:social-defi} zooms in on Defi DAOs, for instance, SushiSwap, Lido, Ribbon, and Olympus DAO, revealing a significant increase in the Gini coefficient over time. This increase signifies a reduction in decentralization due to growing power inequality within these DAOs. It is worth noting that carbonswap, a decentralized exchange built on the security of Uniswap V2 contracts, offers fast and cost-effective DeFi services while also introducing new environmentally sustainable specifications.

On the contrary, certain DAO types such as social and research, exemplified by CityDAO, LexDAO, and Friends with Benefits, exhibit a notably positive decentralization trend, signifying an increase in decentralization over time (Figure~\ref{fig:social-defi2}). It is worth noting that, as indicated in Table~\ref{tab:voting_metric-1}, in the CityDAO system, the issue of biased voter participation rates appears to be more pertinent than biased token weight distribution in terms of decentralization. The entropy is high, while the Gini coefficient for voters' token weight is low. Therefore, our primary focus for analyzing decentralization in CityDAO involves the linear regression results for entropy, as opposed to the Gini coefficient presented in Table~\ref{tab:voting_metric-1}.
A qualitative observation is that CityDAO, for instance, employs quadratic voting, which may have contributed to the improvement in decentralization levels over time. Our findings also show that there are some exceptions of the decentralization level where there is an increase/decrease of decentralization regardless of the DAO type, such as curvedao -- Defi Type, Trufigov -- Defi Type, which has a slight increase trend of decentralization over time (Figure~\ref{fig:social-defi3}). Gitcoin DAO research or funding type and Canbin DAO Social DAO type have a decreasing trend of decentralization over time.

%% file: sections/Discussion.tex
\section{Discussion}

In this section, we discuss the prevailing governance structure of DAOs and the degree to which they realize their envisioned goals as well as \textit{open questions} of DAOs warranting further investigation. Based on the current outcomes of DAOs, we also provide \emph{governance and design implications} for emerging technology governance.

\subsection{DAOs as Complex Socio-Technical Systems} 
DAOs represent a potential empirical framework for complex social computing systems and an opportunity for understanding emerging forms of computational governance~\cite{dupont2017experiments}. 
Through a computational lens, DAOs allow for modeling complex social, organizational, and legal issues as computational challenges~\cite{karp2011understanding, wigderson2019mathematics, zhou2023iterative}, offering new research directions. Prior literature has suggested factors such as token distribution, voter participation, delegation, and voting behavior patterns as influential in DAO governance success. Our findings confirm these hypotheses across a diverse set of DAOs, showing statistically significant correlations between secondary market activities and voting power distribution, thus decentralization.
Specifically, we observe that DAOs focused on social causes or public good (e.g., sports-oriented DAOs) tend to exhibit lower Gini coefficients, reflecting higher decentralization. In contrast, infrastructure, and investment-oriented DAOs demonstrate higher Gini values leading to centralization. Additionally, whale voters in many DAOs, particularly in DeFi and investment categories, are more likely to participate, leading to poor decentralization. Notably, we found a significant power imbalance in DAOs like 1inch, Balancer, dYdX, Fei, GMX, and Gnosis, where low voter participation and high centralization are evident with entropy metrics.

Overall, a trend emerges where most DAOs in the DeFi, investment, and infrastructure sectors show a negative trend in decentralization, with the exception of a few DeFi DAOs, such as Trufigov, which exhibit slight decentralization improvements over time. Conversely, research- and funding-oriented DAOs, such as Gitcoin DAO, and social DAOs like Canbin DAO, show a decrease in decentralization, possibly due to governance contracts and community structures that were beyond the scope of this study. This indicates future research into the specific smart contract governance and off-chain community influencing these trends.

Throughout this study, the results of the Gini index metrics were consistent across both average and historical data points, suggesting that Gini metrics could serve as a useful tool for DAO stakeholders to monitor decentralization in the context of voting power and the influence of governance token design. However, the entropy and Nakamoto metrics were less consistent. While entropy showed some reliability in measuring the impact of voting participation on decentralization, our assessment was limited to active voters, which leaves out the potential centralizing effect of voters with closely aligned interests across elections, often referred to as voting bloc entropy~\cite{austgen2023dao}. This could be further empirically evaluated as a potential metric for decentralization. Another key factor we did not account for was inactive voters, particularly large token holders (whales). It would be valuable to explore the extent to which these inactive voters could influence outcomes if they participated, highlighting the potential importance of the Nakamoto coefficient in measuring the activity of both inactive whales who might disrupt outcomes and inactive grassroots voters who may need encouragement to better represent majority preferences.

\vspace{-2mm}
\subsection{Democratic Infrastructures for Governing Technology} 
Our study highlights key metrics for decentralization, including voting power, grassroots participation, and non-monetary governance tokens. These factors significantly influence the design of decentralized, thus democratic systems aimed at engaging diverse stakeholders in collective decision-making. Traditionally, technology and algorithmic governance have been viewed through institutional or legal frameworks, focusing on regulatory mechanisms like laws and institutions~\cite{denardis2009protocol}, such as research on Internet protocol governance, NIST risk assessment~\cite{maclean2017nist} often centers on technical standards bodies. 
Recent studies have explored ways to involve users in the design and governance of digital technologies, including machine learning algorithms~\cite{smith2020keeping, sharma2024experts}, but these initiatives have not yet achieved widespread adoption. Engineering efforts within DAOs can provide valuable improvements for governance approaches in emerging areas such as AI governance, where the research community is exploring the democratization of AI~\cite{nsf, meta}.

DAOs represent a community-oriented approach to technology governance, typically limited to protocols and contracts on a blockchain. The challenge lies in extending DAO governance to technologies like telecommunications protocols, and AI models development and deployment which are not implemented in a Web3 context. There is potential for addressing sensitive areas such as content moderation in social media or policing through algorithmic oversight, allowing affected parties to participate to govern the technology within their values and norms. This raises the question of how DAOs or similar social technologies can effectively govern a broader range of technologies beyond their current scope. Understanding best practices for DAO governance can help re-imagine and potentially redesign current political processes~\cite{bernholz2021digital}. Considering both DAO technological innovation and theoretical underpinning from political science~\cite{bernholz2021digital}, we can design methods and tooling to engage diverse groups of users in governing AI models to collectively make a decision to shape AI rules whereas the technological components can enable transparent and trustworthy public participation, and minority voice representation.

\section{Conclusion}
Our study found nuanced differences in DAO types, particularly in how the nature of governance tokens (monetary vs. non-monetary) impacts decentralization levels; token-based governance tends to disincentivize grassroots holders in the decision-making. Our work highlights increased grassroots participation and lower variance in voting power within a DAO correlates with a higher level of decentralization. This finding is closely related to core topics in political science, platform governance such as the transfer and allocation of power and one's influence in the community decision-making and the emergence and consequences of different governance systems. We conclude with the advantage of studying the governance of DAOs where it can serve as an interesting testbed for exploring social choice experiments, potentially improving the current governance structures in emerging technology, like AI.  